\documentclass[twocolumn]{aastex61}
\submitjournal{ApJS}

\usepackage{graphicx}
\usepackage{CJK}
\usepackage{amsmath}

\begin{document}
\title{M Subdwarf Research. I. Identification, Modified Classification System, and Sample Construction}

\correspondingauthor{A-Li Luo}
\email{lal@nao.cas.cn}

\author[0000-0003-1454-1636]{Shuo Zhang}
\affiliation{Key Laboratory of Optical Astronomy, National Astronomical Observatories, Chinese Academy of Sciences, Beijing 100012, China}
\affiliation{University of Chinese Academy of Sciences, Beijing 100049, China}

\author[0000-0002-0786-7307]{A-Li Luo}
\affiliation{Key Laboratory of Optical Astronomy, National Astronomical Observatories, Chinese Academy of Sciences, Beijing 100012, China}
\affiliation{University of Chinese Academy of Sciences, Beijing 100049, China}

\author{Georges Comte}
\affiliation{Aix-Marseille Univ, CNRS, CNES, LAM, Laboratoire d'Astrophysique de Marseille, Marseille, France}

\author[0000-0002-8916-1972]{John E. Gizis}
\affiliation{Department of Physics and Astronomy, University of Delaware, Newark, DE 19716, USA}

\author[0000-0001-6767-2395]{Rui Wang}
\affiliation{Key Laboratory of Optical Astronomy, National Astronomical Observatories, Chinese Academy of Sciences, Beijing 100012, China}
\affiliation{University of Chinese Academy of Sciences, Beijing 100049, China}

\author{Yinbi Li}
\affiliation{Key Laboratory of Optical Astronomy, National Astronomical Observatories, Chinese Academy of Sciences, Beijing 100012, China}

\author{Li Qin}
\affiliation{University of Chinese Academy of Sciences, Beijing 100049, China}

\author{Xiao Kong}
\affiliation{Key Laboratory of Optical Astronomy, National Astronomical Observatories, Chinese Academy of Sciences, Beijing 100012, China}
\affiliation{University of Chinese Academy of Sciences, Beijing 100049, China}

\author{Yu Bai}
\affiliation{Key Laboratory of Optical Astronomy, National Astronomical Observatories, Chinese Academy of Sciences, Beijing 100012, China}

\author{Zhenping Yi}
\affiliation{School of Mechanical, Electrical and Information Engineering, Shandong University at Weihai, Weihai 264209, China}

\begin{abstract}

We propose a revision of the system developed by L$\rm\acute{e}$pine et al. for spectroscopic M-subdwarf classification. Based on an analysis of subdwarf spectra and templates from Savcheva et al., we show that the CaH1 feature originally proposed by Gizis is important in selecting reliable cool subdwarf spectra. This index should be used in combination with the [TiO5, CaH2+CaH3] relation provided by L$\rm\acute{e}$pine et al. to avoid misclassification results. In the new system, the dwarf–subdwarf separators are first derived from a sample of more than 80,000 M dwarfs and a ``labeled'' subdwarf subsample, and these objects are all visually identified from their optical spectra. Based on these two samples, we refit the initial [TiO5, CaH1] relation and propose a new [CaOH, CaH1] relation supplementing the [TiO5, CaH1] relation to reduce the impact of uncertainty in flux calibration on classification accuracy. In addition, we recalibrate the $\rm\zeta_{TiO/CaH}$ parameter defined in L$\rm\acute{e}$pine et al. to enable its successful application to Large Sky Area Multi-Object Fiber Spectroscopic Telescope (LAMOST) spectra. Using this new system, we select candidates from LAMOST Data Release 4 and finally identify a set of 2791 new M-subdwarf stars, covering the spectral sequence from type M0 to M7. This sample contains a large number of objects located at low Galactic latitudes, especially in the Galactic anti-center direction, expanding beyond previously published halo- and thick disk-dominated samples. Besides, we detect magnetic activity in 141 objects. We present a catalog for this M-subdwarf sample, including radial velocities, spectral indices and errors, and activity flags, with a compilation of external data (photometric and Gaia Data Release 2 astrometric parameters). The catalog is provided online, and the spectra can be retrieved from the LAMOST Data Release web portal.

\end{abstract}

\section{Introduction} \label{sec:sec1}

M subdwarfs are Galactic fossils with lifetimes much longer than the Hubble time \citep{1997ApJ...491L..51L}. These faint low-mass stars were originally discovered because they combined a large proper motion and a low luminosity \citep{1939ApJ....89..548K} and were subsequently found to share similar kinematics as the inner halo and thick disk stellar populations \citep{1991ApJS...77..417K, 2013AJ....145...40B}. M subdwarfs are rare in the solar neighborhood, but they are supposed to be the largest stellar component of the Milky Way’s halo \citep{2013AJ....145...40B}, while their cousin M main-sequence dwarfs are the most numerous stellar inhabitants of the Milky Way disk \citep{2002AJ....124.2721R, 2003ApJ...586L.133C, 2010AJ....139.2679B}. Studies of M-subdwarf spectra show that they are metal-poor objects compared with the common M dwarfs of near-solar metallicity \citep{1997AJ....113..806G, 2007ApJ...669.1235L}, thus making them crucial touchstones of the star formation and metal enrichment histories of the Milky Way \citep{2016AJ....152..153J}.

Initially, cool subdwarfs were selected from high proper motions catalogs rather than by their spectroscopic features, because the efficiency of photographic spectrographs for faint object is poor in the red spectral range, where their molecular absorption bands are most prominent. Rare examples of spectroscopically identified metal-poor M subdwarfs were provided by \citet{1947ApJ...105...96J}. Then several works suggested that subdwarfs are metal-poor stars associated with halo kinematic population \citep{1976ApJ...207..535M, 1978ApJ...220..935M}. \citet{1980ApJ...240..859A} conducted a spectroscopic survey of high-velocity stars and published the spectra of four targets, which appeared to have ``extreme metal deficiency'' compared with the ``normal'' M subdwarfs. Similar stars were identified spectroscopically in searches for nearby white dwarfs \citep{1979ApJ...233..226L}, for Population II halo stars \citep{1984ApJ...286..269H}, and in a survey of cool M dwarfs \citep{1982PASAu...4..417B}, all focused on faint stars that have high proper motions. Until the end of the 20th century, the term ``subdwarf'' encompassed any number of characteristics including photometric, spectral, or kinematic properties. Meanwhile, a star, whether it shows only one or more of these characteristics, might still be considered as a genuine ``subdwarf'' \citep{1991ApJS...77..417K}.

A spectroscopic analysis is fundamental to access the physical parameters of stars (effective temperature, surface gravity, and surface chemical composition) and their kinematics (radial velocity (RV) component). However, the local scarcity of subdwarfs and their intrinsic faintness for a long time made their spectra difficult to obtain and limited to low or moderate resolution. Thus spectral classification is an important step for understanding the physics of M subdwarfs from low- or moderate-resolution spectra.

The first systematic analysis and classification of M-subdwarf spectra was proposed by \cite{1997AJ....113..806G}, hereafter G97, who divided M dwarfs into three metallicity subclasses (``ordinary'' dM, ``metal-poor subdwarf'' sdM and ``extreme-subdwarf'' esdM) based on the differences between TiO and CaH molecular absorption bands. Empirical relations were drawn from the spectra of a sample of large proper motion or low-luminosity targets. Then \cite{2007ApJ...669.1235L}, hereafter L07, proposed a revision of the system from more than 400 subdwarfs that have halo-like kinematics. L07 defined a metallicity indicator $\rm\zeta_{TiO/CaH}$, which provided a numerical estimate of how the TiO-to-CaH ratio in a star compares to the value measured in solar metallicity objects. Besides, L07 also introduced an additional metallicity subclass of ``ultra-subdwarf'' (usdM) to distinguish the most metal-poor objects and make the classification results consistent with the study of common binaries. In this system, subdwarfs are classified into sdM/esdM/usdM based on their $\rm\zeta$ values.

The classification systems have been used in large spectroscopic data sets from modern spectroscopic surveys, which have fulfilled the task of observing multiple targets simultaneously through mature multi-fiber technology, such as the Sloan Digital Sky Survey \citep[SDSS,][]{2000AJ....120.1579Y} and the Large Sky Area Multi-Object Fiber Spectroscopic Telescope \citep[LAMOST,][]{2012RAA....12.1197C}. The number of spectroscopically identified subdwarfs is thus increasing rapidly \citep[e.g.,][]{2008ApJ...681L..33L, 2014ApJ...794..145S, 2016NewA...44...66Z, 2016RAA....16..107B}. The current largest sample of M-subdwarf spectra \citep{2014ApJ...794..145S}, hereafter S14, contains 3517 targets from SDSS Data Release 7 (DR7).

However, we made a comparison of the G97 system and the L07 system with the S14 subdwarf sample, and the result shows that the L07 system is not identical with the G97 one. Three relations on the [CaH1, TiO5] [CaH2, TiO5] and [CaH3, TiO5] diagrams were first defined in G97 as separators of dM/sdM/esdM, and L07 revised the schemes to a single [CaH2+CaH3, TiO5] relation. Based on a more detailed analysis (which can be seen in Section \ref{subsec:problem}), we suggest that the CaH1 index abandoned by L07 is important when searching for genuine optical spectroscopic subdwarfs (see details in Section \ref{subsec:problem}).

Up until now, the search for M subdwarfs has been limited to high Galactic latitude areas, favoring the detection of halo and thick disk objects. Whether or not a cool subdwarf population could be present in the Galactic thin disk, mixed (or not) with the immense M-dwarf population and having similar (or not) kinematic and dynamic properties, remains an open question. LAMOST data provide an excellent source to address this problem, since this instrument is conducting the most complete spectroscopic survey of the Galactic disk \citep{2015RAA....15.1095L} and is especially targeted to the regions of the Galactic anti-center. Therefore, the need for a reliable classification system well adapted to LAMOST products was felt when searching for spectroscopic M subdwarfs in the LAMOST data set.

\begin{figure*} [ht]
\plotone{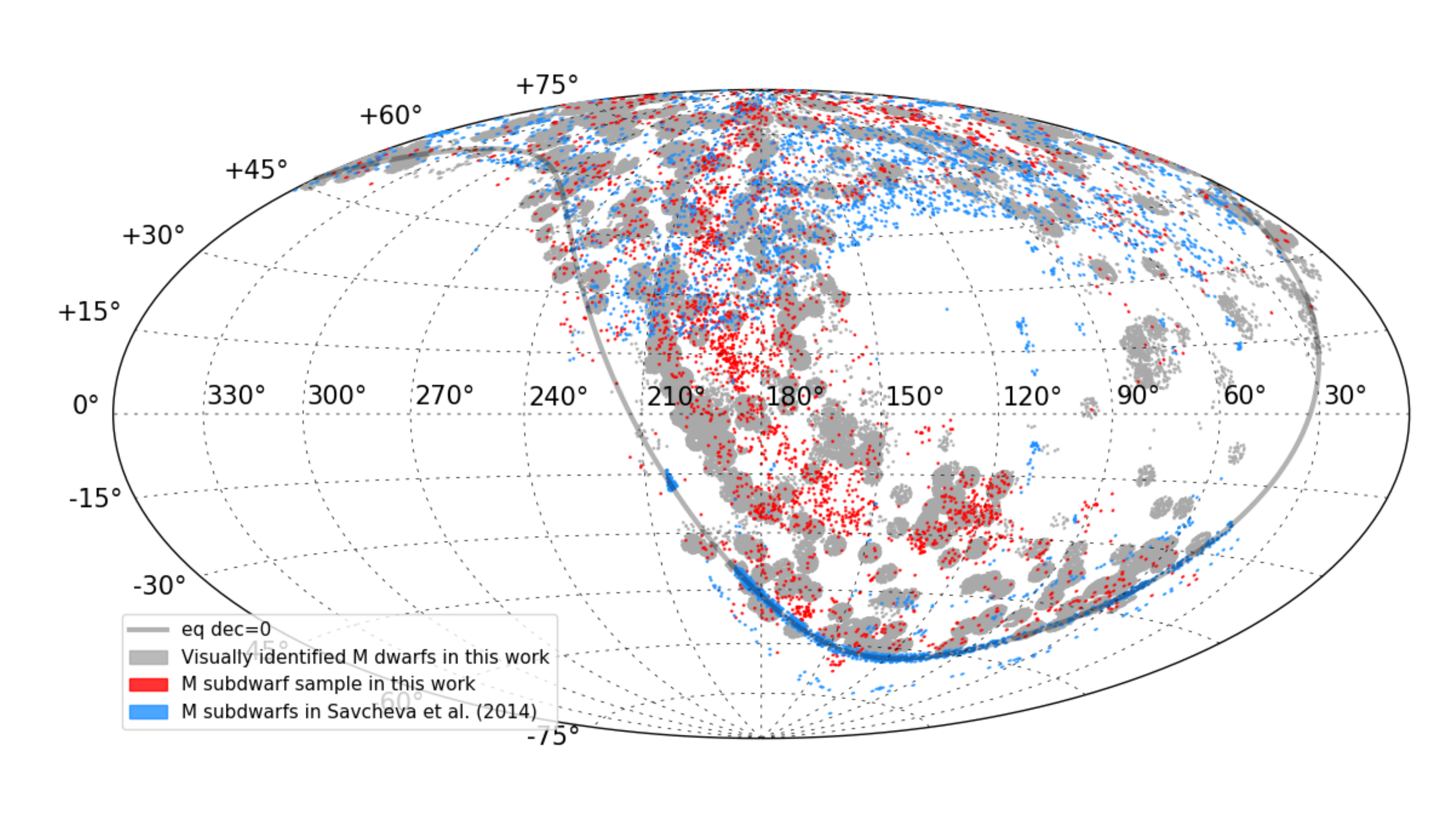}
\caption{The LAMOST DR4 M dwarfs and subdwarfs in the Galactic coordinate system} \label{fig:skymap}
\end{figure*}

In the present work, in order to revise the classification standards, we use more than 90,000 visually identified M-dwarf and M-subdwarf spectra. Then we apply the modified system to the entire LAMOST Data Release 4 (DR4) to select subdwarf candidates. Ultimately, we obtain a sample of 2791 M subdwarfs validated via visual inspection, covering the spectral sequence from M0 to M7. The catalog is available at \url{http://paperdata.china-vo.org/szhang/DR4_Subdwarfs.csv}. Figure \ref{fig:skymap} shows the distribution of these objects in the Galactic coordinate system.

The present work is limited to spectroscopic identification, classification of candidates, and sample selection in LAMOST data. The paper is organized as follows: we review the previous works and propose solutions to existing problems in Section \ref{sec:sec2}. In Section \ref{sec:sec3}, we introduce the LAMOST data set and our RV measurement method. In Section \ref{sec:sec4}, we define the revised classification system, present the final subdwarf sample, and briefly summarize the catalog. Finally, in Section \ref{sec:conclude}, we give a summary and conclusion.

\section{Identification and Spectral Classification of M-subdwarf Stars: A Review of Existing Problems} \label{sec:sec2}

\subsection{Basic Characteristics of M-subdwarf Spectra at Low Resolution}\label{subsec:basic}

\begin{figure}[ht!]
\center
\includegraphics[width=84mm]{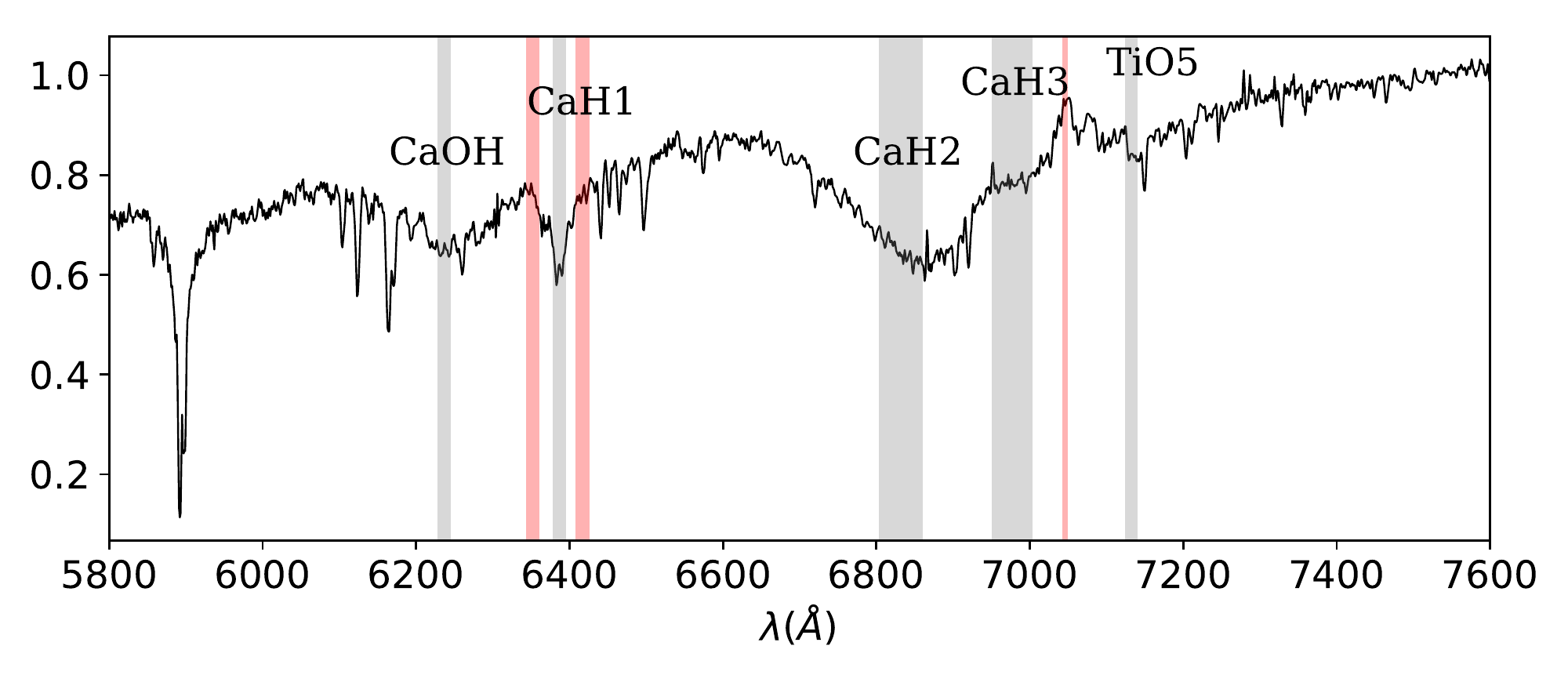}
\caption{Template esdM spectrum in the rest frame. The wavebands used to calculate the spectral indices are marked by the translucent gray bands, corresponding to the ``fea$\_$'' features in Table 1, and the bands marked in red show the ``cont$\_$' features used as pseudo-continuum.}\label{fig:featureband}
\end{figure}

M-subdwarf stars, as cool stars of a low effective temperature \citep[4000$\sim$2400K,][]{2013A&A...556A..15R} have a spectrum dominated by molecular absorption bands, which show considerable overlapping across the whole visible range. The continuum being weak in the blue, the best observing range with standard optical spectrographic instruments is the red and the deep red, where the spectrum is covered by a dense forest of molecular lines, hiding or blending most of the atomic lines used in the usual spectral analysis and diagnostic. At low resolution, molecular absorption bands from metal oxides and hydrides dominate the red range, such as titanium oxide (TiO), vanadium oxide (VO), CaH, and H$_2$O. \cite{1965ApJ...142..925E} tried to identify subdwarfs with MgH and TiO bands, then \cite{1976ApJ...210..402M} proposed a similar method using CaH and TiO bands: because cool subdwarf stars are metal-deficient with respect to normal dwarfs, the metal oxides are depleted with respect to the hydrides at a given effective temperature in their atmosphere, hence the intensity ratio of molecular absorption between an oxide and an hydride should be useful as a separator. The most popular set of indices was defined by \cite{1995AJ....110.1838R} and expanded by \cite{2003ApJ...585L..69L}. We list this classical set of indices in Table 1. Figure \ref{fig:featureband} shows these wavebands on a template spectrum.

\begin{deluxetable*}{cccc}[!ht]
\tablecaption{The features used for classification}\label{tab:features}
\tablecolumns{4}
\tablewidth{0pt}
\tablehead{
\colhead{No.} &
\colhead{Feature} &
\colhead{$\rm\lambda_{Begin}$($\rm\AA$)} &
\colhead{$\rm\lambda_{End}$($\rm\AA$)}
}
\startdata
1 & fea$\_$CaOH & 6230.0 & 6240.0 \\
2 & fea$\_$CaH1 & 6380.0 & 6390.0 \\
3 & fea$\_$CaH2 & 6814.0 & 6846.0 \\
4 & fea$\_$CaH3 & 6960.0 & 6990.0 \\
5 & fea$\_$TiO5 & 7126.0 & 7135.0 \\
6 & cont$\_$CaOH & 6345.0 & 6355.0 \\
7 & cont$\_$CaH1$\_$1 & 6345.0 & 6355.0 \\
8 & cont$\_$CaH1$\_$2 & 6410.0 & 6420.0 \\
9 & cont$\_$CaH2 & 7042.0 & 7046.0 \\
10 & cont$\_$CaH3 & 7042.0 & 7046.0 \\
11 & cont$\_$TiO5 & 7042.0 & 7046.0 \\
\enddata
\center
\tablecomments{The useful quantity is the mean flux across the band, and the bands are in air wavelength.}
\end{deluxetable*}

The value of a spectral index is given by:  
\begin{equation}\label{euqa:index}
\rm Index = \frac{F_{fea}}{F_{cont}}
\end{equation}

where $\rm F_{fea}$ is the mean flux in the molecular absorption feature band, and $\rm F_{cont}$  is a pseudo-continuum flux, i.e., the mean flux in the waveband from $\rm\lambda_{Begin}$ to $\rm\lambda_{End}$ (for CaH1 index, the pseudo-continuum flux is the average flux of two bands). For example, the CaOH index is measured using the mean flux within fea$\_$CaOH divided by the one within cont$\_$CaOH.

\subsection{Review of the Selection Schemes and Classification Systems}\label{subsec:review}
\subsubsection{G97 system}\label{subsubsec:Gizis}

The spectroscopic classification system of M subdwarfs was initially defined by G97, from spectra of 79 targets mostly selected from large proper motion catalogs, such as Luyten Half-Second Catalogue \citep[LHS,][]{1979lccs.book.....L} and the Lowell Proper Motion Survey \citep{1971lpms.book.....G}. Based on quantitative measures of TiO and CaH features (CaH1, CaH2, CaH3, and TiO5), G97 defined a set of empirical relations on the CaH$\rm n-$TiO5 ($\rm n$=1,2,3) diagrams that classified M-dwarf spectra into three subclasses corresponding to increasing metal-poor levels: dwarfs of solar metallicity (dM), normal subdwarfs (sdM), and extreme subdwarfs (esdM).

The initial condition proposed by G97 to identify a spectroscopic subdwarf was the [CaH1, TiO5] relation:
\begin{equation} \label{equa:g97}
\begin{split}
\rm CaH1 &\rm < 0.695\times \rm TiO5^3\ -\ 0.818\times \rm TiO5^2\\
&\rm +\ 0.413\times \rm TiO5\ +\ 0.651
\end{split}
\end{equation}

G97 suggested it was the primary dwarf/subdwarf separator and added a [CaH2, TiO5] relation to separate classic subdwarfs from extreme subdwarfs. Finally, G97 assigned the spectral subclass (linked to the effective temperature) using the CaH3 index. For the coolest stars with TiO5$<$0.49, G97 also suggested that the CaH2 or CaH3 index must be used to avoid inaccuracy caused by the ``saturation'' of the [CaH1, TiO5] relation.

\subsubsection{L07 system}\label{subsubsec:Lepine}

\cite{2003AJ....125.1598L} later suggested that the [CaH1, TiO5] relation could be abandoned due to its shorter dynamic range and the lower signal to noise of CaH1 band in very cool stars. They proposed its replacement by a [CaH2+CaH3, TiO5] relation, as producing a result almost equivalent to the former classification for the same stars, according to \cite{2003ApJ...585L..69L}. Along this way, \cite{2006ApJ...645.1485B} went one step further, defined two separators on the [CaH2+CaH3, TiO5] diagram to differentiate between dM/sdM/esdM:

\begin{equation} \label{equa:B06-1}
\begin{split}
\rm sdM: &\rm (CaH2+CaH3) <\ 1.31\times TiO5^3\\\
&\rm -\ 2.37\times \rm TiO5^2\ +\ 2.66\times \rm TiO5\ -\ 0.20
\end{split}
\end{equation}
\begin{equation} \label{equa:B06-2}
\begin{split}
\rm esdM: &\rm (CaH2+CaH3) <\ 3.54\times \rm TiO5^3\\\
&\rm -\ 5.94\times \rm TiO5^2\ +\ 5.18\times \rm TiO5\ -\ 1.03
\end{split}
\end{equation}

From a much larger published sample of M-type spectra selected from the high proper motion catalog L$\rm acute{e}$pine and Shara Proper Motion (LSPM)-north \citep{2005AJ....129.1483L}, L07 proposed an update of this classification system based on an empirical calibration of the TiO/CaH ratio for stars of near-solar metallicities. For this, they introduced a parameter $\rm\zeta_{TiO/CaH}$, which quantifies the weakening of the TiO band-strength due to the metallicity effect, with values ranging from $\rm\zeta=1$ for stars of near-solar metallicities to $\rm\zeta=0$ for the most metal-poor (and TiO depleted) subdwarfs. The $\rm\zeta$ parameter is defined as

\indent
\begin{equation}\label{equa:zeta}
\rm \zeta_{CaH/TiO}=\frac{1-TiO5}{1-[TiO5]_{Z_\odot}}
\end{equation}
the $\rm [TiO5]_{Z_\odot}$ is a third order polynomial fit of the TiO5 spectral index as a function of the CaH2+CaH3 index. In L07, these indices are measured in spectra of a kinematical selection of local disk dwarfs of roughly the solar metallicity, giving
\begin{equation} \label{equa:cah23-tio5}
\begin{split}
\rm [TiO5]_{Z_\odot}\ &=\ -0.05\\
&\rm -0.118\times (CaH2\ +\ CaH3)\\
&\rm +0.670\times (CaH2\ +\ CaH3)^2\\
&\rm -0.164\times (CaH2\ +\ CaH3)^3
\end{split}
\end{equation} 

L07 suggested that the separation between subdwarfs and ordinary dwarfs could be effective when applying the single condition: $\rm\zeta<0.825$. L07 also refined the scheme by introducing an additional class of usdM corresponding to the most metal-poor ones.

Since then, L07 system has been widely used in the identification, classification, and subtype determination of the cool subdwarf spectra \citep[e.g.,][]{2008ApJ...681L..33L, 2013AJ....145...40B, 2014ApJ...794..145S, 2016RAA....16..107B, 2017MNRAS.464.3040Z}.

\subsubsection{$\rm\zeta$ and the Metallicity}\label{subsubsec:metallicity}

There is a major problem with the sdM/esdM/usdM subclasses in the classification systems above. The subclass defined by CaH and TiO indices were originally used to infer metallicity levels, although the separators do not seem to run parallel to the lines isometallicity drawn from synthetic model grids, such as the NextGen grid used by L07. \cite{2008AJ....136..840J} then compared the Gaia stellar atmosphere model grid with 88 subdwarf spectra and pointed out that these indices are affected in complicated ways by combinations of the temperatures, metallicities, and gravities of subdwarfs. \cite{2013AJ....145..102L} also compared observed spectra with BT-Settl stellar atmosphere model grids and concluded that the TiO/CaH ratio is not primarily sensitive to the classical metallicity value [Fe/H], but rather depends on the [$\rm\alpha$/H] because O, Ca, and Ti are all $\rm\alpha$-elements. Variations in [$\rm\alpha$/Fe] would thus weaken the correlation between $\rm\zeta$ and [Fe/H]. More information is necessary to assess stellar atmospheric parameters of a subdwarf, such as a binary membership or detailed model fitting of a high-resolution spectrum \citep{2016A&A...596A..33R}.

Yet even so, the separators defined by L07 in the [CaH2 +CaH3, TiO5] space can still indicate a rough ordering in the metallicity, as they were verified on a sample of resolved subdwarf binaries, and in fact the metallicity subclasses are more associated with the kinematics of the objects as they were first determined by subdwarfs with different kinematic properties. To separate subdwarf spectra from normal main-sequence dwarfs, the parameter $\rm\zeta_{TiO/CaH}$ still appears to be a robust indicator.

\subsection{Problems in the Dwarf/Subdwarf Separation and Subclass Limits}\label{subsec:problem}

\subsubsection{Specific Problems with S14 Sample Selection}\label{subsubsec:s14_rv}

\begin{figure*}[!ht]
\plotone{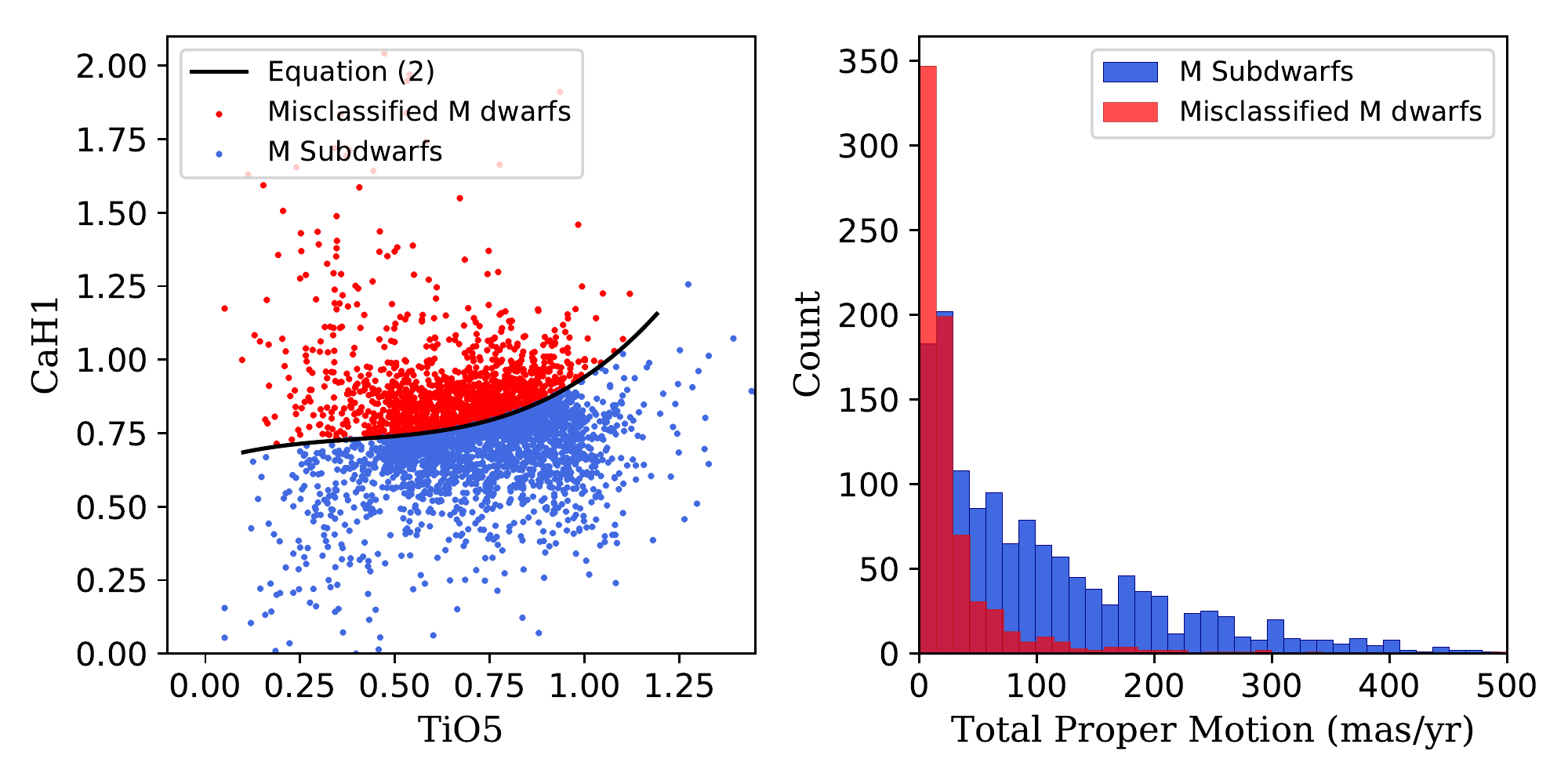}
\caption{Distribution of the subdwarfs from the  \cite{2014ApJ...794..145S} catalog in the [CaH1, TiO5] index diagram and histogram of their proper motions. The subdwarfs were selected by a [CaH2+CaH3]-TiO5 relation defined in \cite{2007ApJ...669.1235L} from SDSS DR7. The black curve on the left diagram is the original subdwarf selection criterion, i.e., the CaH1–TiO5 relation defined in \cite{1997AJ....113..806G} and abandoned by \cite{2007ApJ...669.1235L} There are 1447 targets (red dots as group I) in this ``subdwarf'' sample that do not fulfilling the CaH1-TiO5 relation, which thus might be misclassified M dwarfs. The other ones, which could be considered as reliable subdwarfs, are figured as blue dots (group II). The right panel shows that most group I stars do not exhibit large proper motions.}
\label{fig:sav_cah1}
\end{figure*}

To search for subdwarfs, S14 adopted the L07 system as a filter to select candidates from M-type spectra in SDSS DR7 and finally identified 3517 M subdwarfs with optical spectra. However, two problems may be identified in this subdwarf sample:

(1) As shown in Figure \ref{fig:sav_cah1}, a large fraction (1447) of objects in the sample do not seem to satisfy Equation (\ref{equa:g97}), which means the L07 system is not identical to the original G97 system. In fact, these targets do not exhibit classical ``subdwarf'' characteristics, such as large proper motions. This leads to the suspect that there is a group of stellar objects that have subdwarf-like [CaH2+CaH3]-TiO5 relationship but which might not be classified as genuine ``subdwarf''. Hence, neglecting the primary condition defined by G97--the CaH1 feature--leads to the misclassification results of some stellar objects when using only the L07 system.

\begin{figure*}[!ht]
\plotone{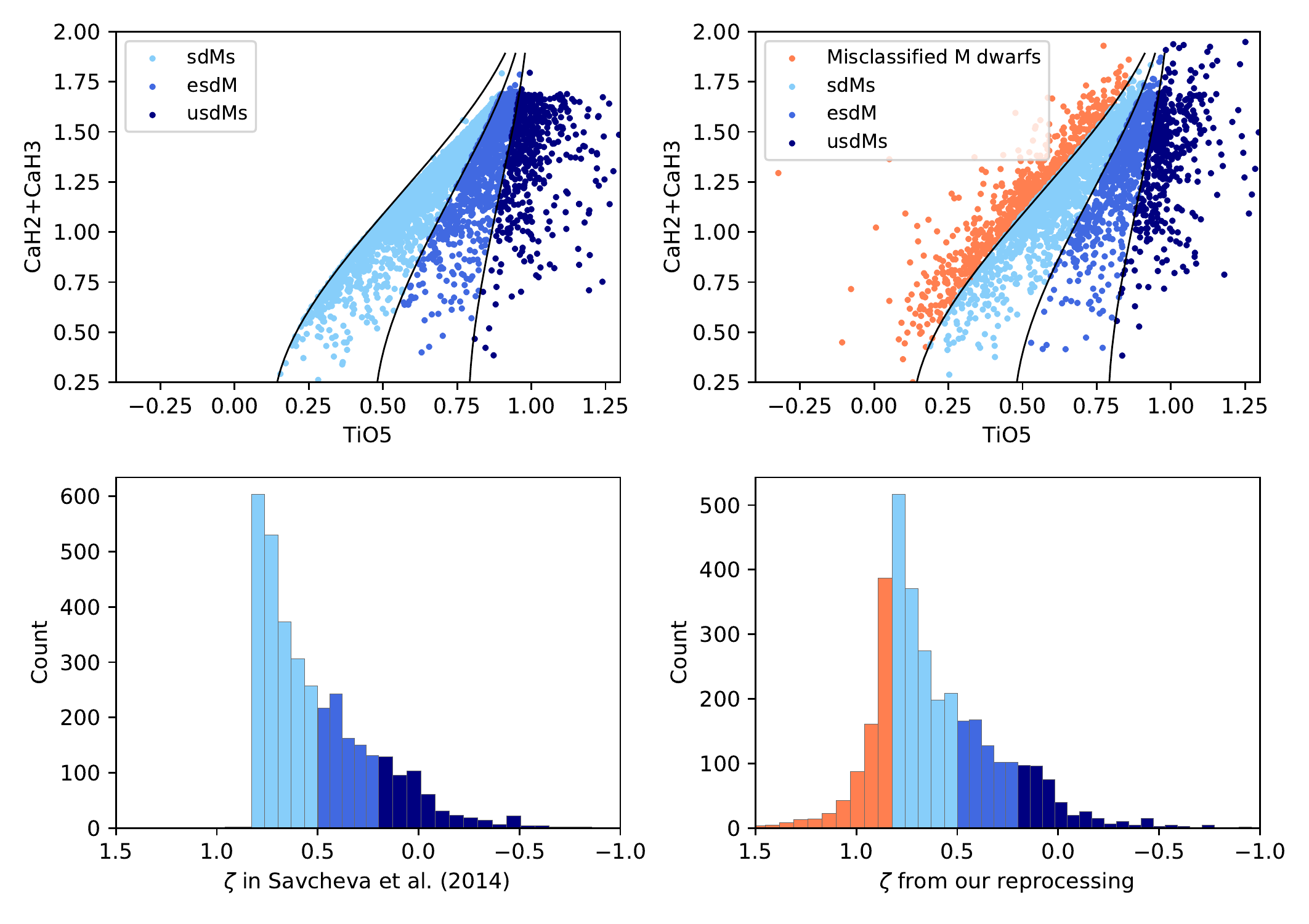}
\caption{Sensitivity of the spectral classification indicators to the spectrum rest-frame reduction accuracy. The two left panels show the distribution of spectral indices and $\zeta$ parameter taken from the large subdwarf catalog of \cite{2014ApJ...794..145S}. They are based on spectra reduced with RVs from \cite{2011AJ....141...97W}, obtained by cross-correlation with ordinary M-dwarf templates. The subdwarfs are selected from dwarfs using $\zeta<0.825$ and were further divided into sdM (light blue), esdM (blue), and usdM (dark blue) based on decreasing $\zeta$ values. The right panels show our reprocessing of the S14 sample using the set of RVs (also provided by S14, but not used by them) based on cross-correlation with purposely built subdwarf templates. The orange dots represent 744 objects that escape the selection criterion and have $\zeta>0.825$, which makes them probable misclassified ordinary dwarfs.} 
\label{fig:sav_RV}
\end{figure*}

(2) In the work of S14, the online catalog provided CaH, TiO indices and $\rm\zeta$ values for the M-subdwarf sample, a large part of which were directly taken from \cite{2011AJ....141...97W}. These measurements were performed on spectra reduced to the rest frame using RV measurements by cross-correlation with M-dwarf templates. As mentioned above, S14 adopted $\rm\zeta<0.825$ as the subdwarf selection criterion. However, S14 also provided a value of RV for each target derived from cross-correlation with a set of M-subdwarf template spectra built by the authors. We have recomputed the CaH, TiO spectral indices and the $\rm\zeta$ index using these last RV values: the results plotted in Figure \ref{fig:sav_RV} show that there are 744 objects with $\rm\zeta>0.825$, which should be classified as dwarfs rather than subdwarfs according to this selection criterion.

These discrepancies clearly arise from the accuracy of the rest-frame reduction of spectra used to measure indices. The whole process is highly sensitive to the template spectra used in the cross-correlation. Therefore, a method independent of templates to measure RVs of the candidates is mandatory for us to ensure the accuracy of the final sample, since the LAMOST pipeline does not contain subdwarf templates.

\subsubsection{The Need to Recalibrate $\zeta_{TiO/CaH}$}\label{subsubsec:recal}

\cite{2013AJ....145..102L} made a comparison of spectra from the same stars obtained at different observatories, which revealed that spectral band index measurements are dependent on spectral resolution, spectrophotometric calibration, and other instrumental factors. Different spectral resolution, spectrophotometric calibration, and other instrumental factors would all lead to different spectral index values for the same target spectrum \citep{2013AJ....145..102L}, and different data sets may also have their own biases due to selection effects. The authors thus suggested that a consistent classification scheme requires that spectral indices be calibrated and corrected for each observatory/ instrument combination used. For example, \cite{2012AJ....143...67D} recalibrated the definition of $\zeta$ with a sample of SDSS binary systems to correct the bias in $\zeta$ observed in early-type M dwarfs. Further on, \cite{2013AJ....145..102L} recalibrated the $\zeta$ parameter with corrected spectral index values from their new larger dwarf sample. Table \ref{tab:coe} lists the coefficients of the polynomial fit of the [CaH2+CaH3] versus TiO5 mean relation for ordinary dwarfs obtained by these authors.

Hence, we have decided to derive a new calibration of the $\zeta$ index based on LAMOST data.

\subsubsection{Difficulties in Using LAMOST Spectra}

Due to the low luminosity of low-mass stars and efficiency limits of the LAMOST instrument, a large fraction of the M-type stars do not have high-quality spectra. In addition, since the spectral indices are defined as flux ratios, the accuracy of them used for selection and classification may also be affected by continuum slope deformation from flux calibration problems \citep{2016ApJS..227...27D}. For example, the waveband of CaH1 feature is $\rm\sim700\rm \AA$ away from TiO5, hence the inaccuracy of continuum slope may affect the comparison result of the CaH1 index with the TiO5 index.

\cite{2016RAA....16..107B} made an effort to search for M subdwarfs in Data Release 2 (DR2) of LAMOST survey (which contains more than 200,000 M-type spectra). The authors used visual inspection as the final identification tool to avoid mistakes caused by the factors mentioned above. As a result, they finally verified 108 objects from the candidates after a prior selection by G97/L07 schemes. This number very probably underestimates the subdwarf fraction potentially contained in DR2 because the classification standards were defined on much smaller samples and should not be directly applied to LAMOST data.

\subsection{A Short Introduction of Proposed Solutions}\label{subsec:solution}

Taking factors above into account, we conclude that a reliable M-dwarf sample and a carefully verified M-subdwarf sample from LAMOST data must first be built in order to define the separators between subdwarfs and dwarfs. An accurate rest-frame reduction, based on an RV measurement method independent of template spectra is necessary to get correct spectral indices. The dwarfs can be used to calibrate the $\zeta$ parameter, and the CaH1 index should be taken into consideration. In addition, the CaOH index is much closer to CaH1 ($\rm\sim 150 \rm\AA$) than TiO5, and its comparison with CaH1 is as useful as TiO5 in separating subdwarfs from dwarfs. Then, combining revised G97 and L07 schemes, an automated search for subdwarf candidates in the entire LAMOST DR4 can be made. Finally, visual inspection will be adopted as the ultimate tool for confirmation of subdwarf identification.

\section{LAMOST Observation and Data Preparation} \label{sec:sec3} 

\subsection{LAMOST Data Release 4}\label{subsec:lamost}
LAMOST is an all-reflective Schmidt-type telescope located in the Xinglong Station of the National Astronomical Observatory, China (105$^\circ$ E, 40$^\circ$ N). It has a 6.67 m spherical primary mirror, which, combined with the corrector, provides an effective aperture of 4 m. It offers both a large field of view (5$^\circ$) and a large aperture ratio. Feeding 16 double channel spectrographs are 4000 fibers mounted on the focal plane, which allow a high spectral acquisition rate \citep{2012RAA....12.1197C}. As it is dedicated to a spectral survey of celestial objects over the entire available northern sky, both Galactic and extragalactic surveys are conducted. The Galactic one, the LAMOST Experiment for Galactic Understanding and Exploration (LEGUE), focuses on the Galactic anti-center direction, the disk at selected longitudes away from the Galactic anti-center, and the halo \citep{2015RAA....15.1095L}.

Standard techniques are used in the analysis of the spectra. On the original charged coupled device (CCD) image, 4000 pixels are used to record blue and red wavelength regions across 3700-5900 $\rm{\AA}$  and 5700-9000 $\rm{\AA}$, respectively. The blue and red channels are combined, and each spectrum is re-binned to calculate the RV. A combined spectrum is resampled at a scale of 69 km s$^{-1}$ per pixel, i.e., the difference between two adjacent points in the wavelength is $\Delta$log($\lambda$) = 0.0001. The raw CCD data are reduced by the LAMOST data reduction software named LAMOST 2D pipeline. The 2D pipeline performs the tasks of subtraction of dark and bias, correction of flat field, extraction of spectra, subtraction of skylight, calibration of the wavelength, merging of sub-exposures with cosmic-ray elimination, and combination of blue and red wavelength ranges \citep{2015RAA....15.1095L}.

The Data Release 4 of LAMOST regular survey contains 7620,612 spectra covering the entire optical band ($\sim$3700-9000 $\rm{\AA}$) at resolution R$\sim$1800, including 6944,971 stellar spectra, 153,348 extragalactic spectra (galaxies and QSO), and 522,293 spectra classified as Unknown (most of the latter because of insufficient signal-to-noise).

\subsection{RV Measurement}\label{subsec:rv} 

The heliocentric RV is the projection of the star’s 3D velocity along the line of sight and can be measured from the Doppler shift of spectral lines. A conventional and accurate way to determine RV is by cross-correlating the target spectrum with a rest-frame template spectrum of a similar type \citep{1979AJ.....84.1511T}. The LAMOST 1D pipeline provides a redshift measurement ``z''  for each object by matching it with a template spectrum. However, up until now, M-subdwarf templates were not yet available for the 1D pipeline, and thus the RVs measured using ordinary M-dwarf templates for subdwarf candidates would be of insufficient accuracy.

To overcome this problem, we have adopted an alternate RV determination method independent of templates. The natural way is to fit single Gaussian profiles to a set of absorption lines of neutral metals \citep{2014ApJ...794..145S,2016RAA....16..107B}. However, M subdwarfs are basically faint red objects whose spectra often have a low or very low signal-to-noise ratio (S/N), and the useful metal lines at low resolution often do not exhibit nice Gaussian profiles. In many cases, the profiles are constructed by only 3$\sim$5 flux points and are not even symmetric. Therefore, we designed the following process to estimate RV from the Doppler shift of eight prominent absorption lines listed in Table 2. The method is based on the fact that the wavelengths of the eight lines would shift simultaneously due to the Doppler effect caused by radial motion, hence the offsets of the line centers would be identical in the logarithmic wavelength. The details of this method are described below:

\begin{deluxetable}{ccc}
\tablecaption{Absorption Lines Used For RV Measurment}\label{tab:rvlines}
\tablecolumns{3}
\tablewidth{0pt}
\tablehead{
\colhead{Line} & \colhead{Center ($\rm{\AA}$ )}
}
\startdata
K I & 7667.0089 \\
K I & 7701.0825\\
Na I & 8185.5054 \\
Na I & 8197.0766\\
Ti I & 8437.2600 \\
Ca II & 8500.3600 \\
Ca II & 8544.4400 \\
Ca II & 8644.5200 \\
\enddata
\center
\tablecomments{The line centers adopted from NIST are in vacuum wavelength.}
\end{deluxetable}

\begin{enumerate}
\item The wavelengths of a LAMOST spectrum are recorded in exponential form in the fits file header as
\begin{center}
$\lambda = 10^{(head + n*step)}$
\end{center}
where $\emph{head}$ corresponds to the COEFF0 field which records the decimal logarithm of the central wavelength of the first pixel, (usually around 3.5682), $n=0,1,2,3,...,$ and step$\equiv$ $0.0001.$ Therefore the observational accuracy of RV when the shift of a line is defined to 1 step unit i.e., when $\rm\Delta n=\pm1$, is $\rm c*(1-10^{\rm head+\Delta n*0.0001}) \approx \pm69$ km s$^{-1}$. We recall that this corresponds to the sampling step of the pipeline reconstructed spectrum: the actual physical spectral resolution (somewhat variable along the spectral range) is of the order of 2.5 sampling steps, but naturally, radial velocity measurement methods actually allow final accuracies of a fraction of a spectrum pixel.
\item Since the LAMOST subdwarfs are within 1 kpc of the Sun (from apparent magnitude considerations), the Sun's escape velocity---550 km s$^{-1}$ \citep{2014ApJ...794...59K}---can be used as a proxy for the escape velocity of the entire sample. When $\rm \Delta n=\pm10$ the corresponding RV is $\pm$690 km s$^{-1}$, hence we suppose that the shift of a line would be less than 10 step units.
\item Without any prior knowledge of the type of spectrum (late K star, M star, QSO or Unknown), for each line in Table 2, we get the wavelength step index $n_0$ in the spectrum corresponding to its rest-frame wavelength, and then find the $\Delta n$ of the lowest flux point in the range [$\lambda_{n_0-10}, \lambda_{n_0+10}$]. We insist that this is an entirely automated process that could provide positions of ``true'' absorption lines as well as simple accidental deeps in the object's continuum, but its  virtue is that the ``line'' profiles themselves are not required to be gaussian.  
\item We count the number of ``lines'' affected by each $\Delta n\in[-10, 10]$. The $\Delta n$ corresponding to the largest number of lines is adopted as our target $\Delta n*$. Note that the number of lines should be at least 3 to avoid a large random error, otherwise we'll flag the target as unable to be measured. And at later visual inspection stage of the selection process, we also filter out the objects with more than 3 ``lines'' which are merely local flux minima instead of real absorption lines. 
\item The ``lines'' affected by the shift $\Delta n*$ are then used to measure the RV: this is achieved on an interpolated spectrum with artificially 15 times increased resolution to acquire a better precision \citep{2007AJ....133..531B}. The mean value of the RVs obtained from these lines is our final RV, and the standard deviation is the error. 
\end{enumerate}

\begin{figure}
\center
\includegraphics[width=84mm]{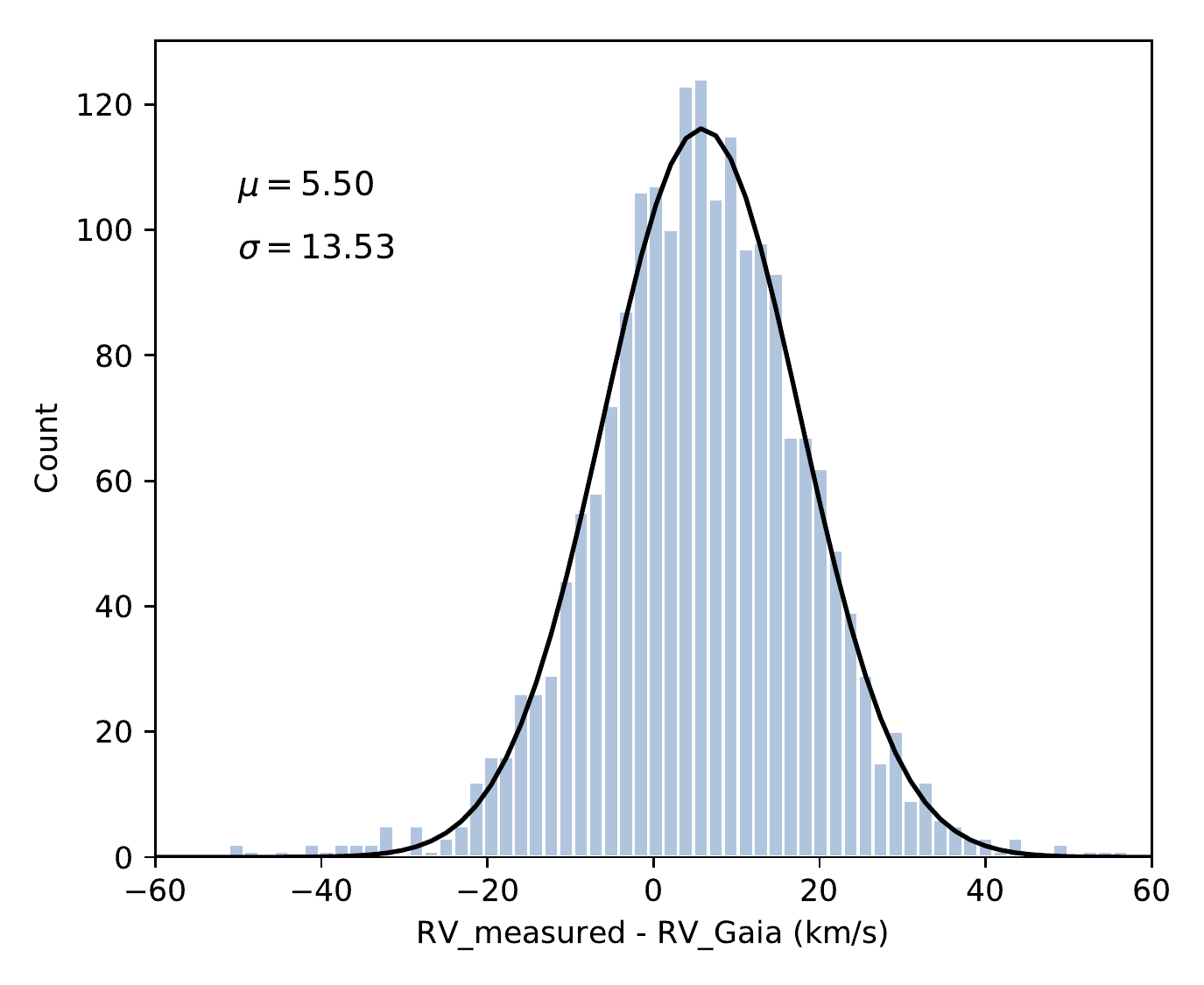}
\caption{Comparison of RVs measured in this work (Subsection \ref{subsec:rv}) on 1926 stars in the M dwarf comparison sample with RVs provided by Gaia radial velocity survey. The Gaia spectrometer collects medium resolution (R $\sim$ 11,700) spectra across the wavelength range 845-872 nm centered on the Calcium triplet region \citep{2018A&A...616A...5C}.} \label{fig:rv_err}
\end{figure}

Our RV measurements have a median achieved typical uncertainty of 13.9 km s$^{-1}$. In Figure \ref{fig:rv_err} we compare the RVs of 1926 M dwarfs from our M dwarf sample (see Section 4.1 below) with the RVs provided by Gaia DR2 on the same stars. Our measurements show a fair agreement with Gaia RVs with a offset of 5.50 km s$^{-1}$ and a dispersion of 13.5  km s$^{-1}$. Similar systematic offsets are also found in other comparisons: 4.54 km s$^{-1}$ between LAMOST and APOGEE \citep{2018arXiv180707625A}, and 3.8 km s$^{-1}$ between LAMOST and MMT+Hectospec \citep{2015MNRAS.449..162H}.
Note that four of the eight absorption lines (the K I doublet and the Na I doublet) are located within telluric absorption bands and sometimes the line centers are altered by insufficiently corrected telluric absorption by O$_2$ and H$_2$O. To solve this problem, we assign a same weight to each of the eight lines of Table 2 and impose that three ``lines'' at least must be detected to provide a valuable result. In addition, we also inspect the 7000-9000 $\rm{\AA}$ range of each spectrum in our final subdwarf sample and remove the objects which do not show identifiable absorption lines by eye-check to improve the reliability of the final results.

\section{Search for M subdwarfs in LAMOST DR4} \label{sec:sec4} 

In order to search for M-subdwarf spectra in the entire LAMOST DR4, we modify the L07 system and apply its new version onto the data set. The implementation process to obtain the final subdwarf sample is thus as follows:
\begin{enumerate}
\item Selection of M-dwarf comparison sample and of the subdwarf labeled sample. We visually inspect the entire M-type stellar spectra collection (99,671) from LAMOST fourth year regular survey. Among this sample, 325 subdwarfs and 90,018 ordinary M dwarfs have been identified. Adding the subdwarfs from \cite{2016RAA....16..107B}, the identified M-subdwarf sample contains more than 400 objects. We call this subsample the subdwarf ``labeled'' sample. The inspection process is discussed in Section \ref{subsec:first_sample}.
\item Calibration of $\zeta$. Based on the M-dwarf sample, we recalibrate the $\zeta$ index as initially defined in L07 with the fitting of the [TiO5, CaH2+CaH3] relation. We adopt as new maximal $\zeta$ values 0.75, 0.5, and 0.2 to separate M-type subdwarf stars into sdM/esdM/usdM, respectively. The details are described in Section \ref{subsec:cal_zeta}.
\item Definition of CaH1 separators. We refit the original G97 [TiO5, CaH1] relation using LAMOST subdwarf ``labeled'' sample and derive Equation (\ref{equa:cah1_tio5}). We also define a new [CaOH, CaH1] relation (Equation (\ref{equa:cah1_caoh})) for supplement. The equations are shown in Figure \ref{fig:cah1_equas}. The process is detailed in Section \ref{subsec:new_cah1}.
\item Determination of the spectral subtype using Equation (\ref{equa:spt}).
\item Searching for M subdwarfs in LAMOST DR4. Applying the constraints above on 1271,426 spectra from DR4 data set, we obtain nearly 10,000 candidates. Finally, 2791 targets among these candidates are confirmed as M subdwarfs by visual inspection. Section \ref{subsec:final_sample} gives the details and Section \ref{subsec:catalog} introduces the subdwarf catalog and the additional external data.
\end{enumerate}

\subsection{The Subdwarf ``Labeled'' Sample and the M-dwarf Comparison Sample} \label{subsec:first_sample} 

The initial step consists in assembling two reference samples from LAMOST data: one of ordinary M dwarfs, the other of carefully validated M subdwarfs. These allow the revision of the various selection and classification tools of G97 and L07 in order to apply them to LAMOST spectra. Since the existing standard M-subdwarf spectral templates are not perfectly appropriate for LAMOST spectra, we choose the most reliable identification/confirmation method underlined in all the former works--visual inspection--as our first identification tool to assemble a reliable subdwarf sample from LAMOST data. Although \cite{2016RAA....16..107B} provided 108 verified M subdwarfs from LAMOST DR2, the candidates submitted to visual inspection were provided by screening through the G97/L07 systems. Therefore, to obtain a larger subdwarf sample, and while we select a comparison sample of purely ordinary M dwarfs, the visual inspection was targeted at the entire M-type spectra data set (auto-classified by LAMOST 1D pipeline) from the fourth year of the LAMOST regular survey, containing nearly 100,000 spectra. In this process, potential contaminant,s such as M giants, double stars, and unrecognizable objects, can be removed, and the spectral subtype of each M-dwarf spectrum can also be verified based on the eye-check result.

To achieve this arduous task, we use the manual ``eyecheck'' mode of the Hammer spectral typing facility \citep{2007AJ....134.2398C}, an IDL code that uses the relative strength of a series of spectral features in the range of 4000-9100 $\rm{\AA}$ for classifying stellar spectra \citep{2008AJ....136.2022L,2011AJ....141...97W,2012AJ....143...67D,2014ApJ...794..145S}. \cite{2014AJ....147...33Y} have modified the Hammer code to better adapt it to LAMOST M spectra. The manual mode of Hammer allows the user to compare a target spectrum with a collection of M-dwarf template spectra and to assign a label to each target.

Referring to the aspect of single subdwarf spectra and/or templates provided in previous works, such as G97, L07, \cite{2008AJ....136..840J}, S14, \cite{2016RAA....16..107B}, we base our initial assignment of a target spectrum into the ``giant'', ``dwarf'' or  ``subdwarf'' category on the following basic recipes:

(1) The K I doublet around 7700 $\rm{\AA}$ and Na I doublet around 8200 $\rm{\AA}$ are highly gravity-sensitive, thus they are almost invisible in giants, but are prominent both in dwarfs and subdwarfs. Compared with dwarfs, the CaH1 absorption band of an M-giant spectrum of the same spectral type is almost invisible, as is the CaH3 band, while its TiO5 minimum is as deep as the CaH2 minimum.

(2) The most obvious differences between M subdwarfs and dwarfs are the features around 6200–6400 $\rm{\AA}$ and 6800–7200 $\rm{\AA}$. CaH1 absorption band is much more prominent in subdwarf spectra than in dwarfs, which is almost as deep as CaOH and even deeper, respectively, for esdM and usdM. Regarding subdwarfs, the TiO5 minimum is less deep than the CaH2 minimum for sdM, is even less deep than the CaH3 minimum for esdM, and the entire TiO absorption band is almost invisible for usdM.

Note that the reddest part of TiO5 is blended with a strong atomic line 7148 $\rm{\AA}$ of Ca I, which is very close to the third (and deepest in normal conditions) ``tooth''-like absorption of the TiO, which makes it easy to misidentify the minimum of TiO5 on low-resolution noisy spectra.

In this step, we label each target with a spectral subtype (M0, M1, KM9) and a type information, such as ``dwarf'', ``giant'', ``subdwarf'', ``double star'', or ``odd''. As a result, 325 subdwarfs are selected and validated. Combined with the 108 M subdwarfs from \cite{2016RAA....16..107B}, we now have a ``labeled'' subdwarf sample that includes 433 visually identified spectra in total.

Besides, 90,018 ordinary M dwarfs are identified, of which 83,213 have measurable RVs: these define our M-dwarf comparison sample. This comparison sample is also used for the calibration of $\zeta$.

\subsection{Calibration of $\rm\zeta_{TiO/CaH}$} \label{subsec:cal_zeta}

\begin{figure}
\plotone{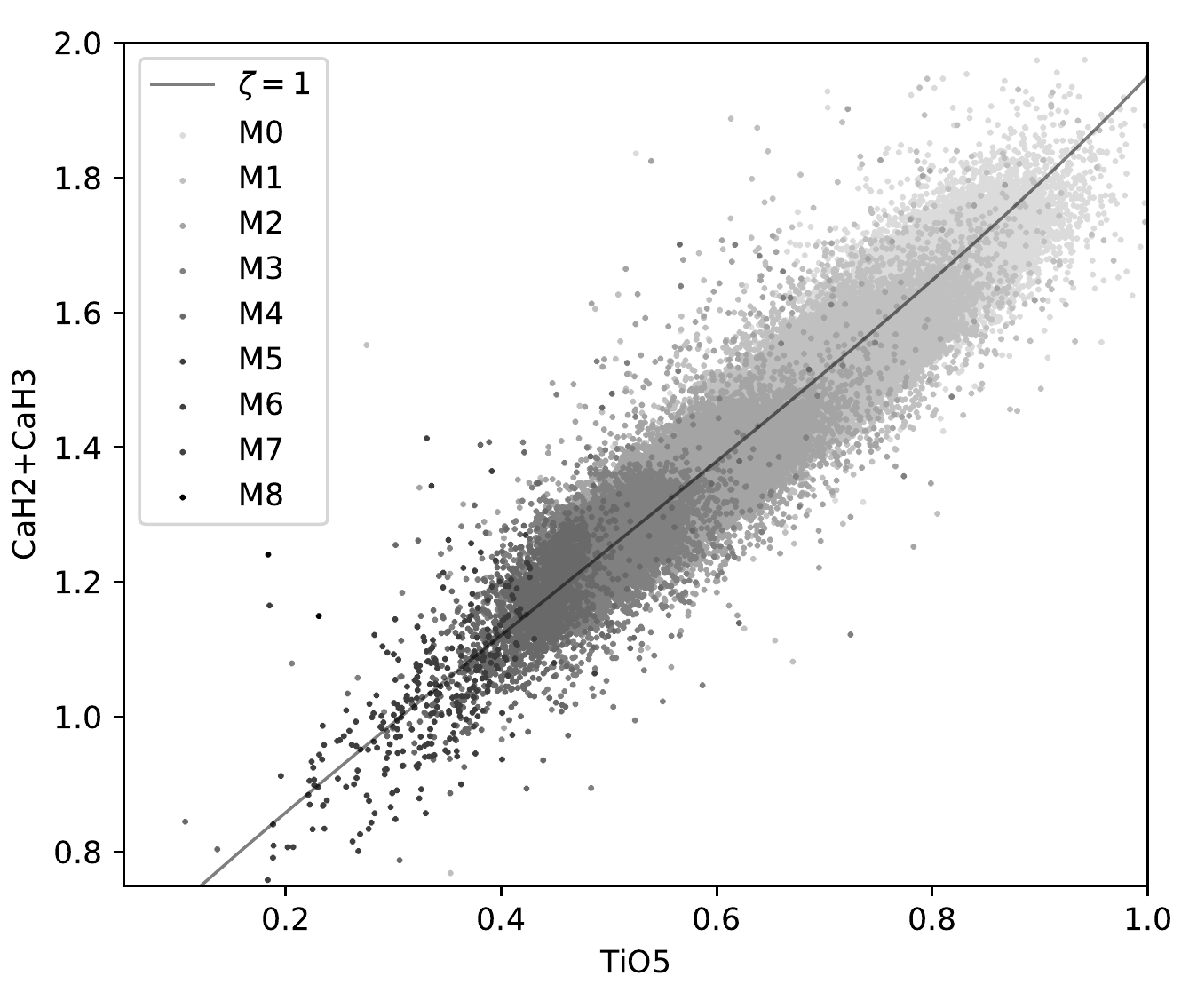}
\caption{Recalibration of the $\rm\zeta_{TiO/CaH}$ parameter (denoted as $\rm\zeta$ for short) based on a sample of 83,213 M-dwarf spectra from the LAMOST regular survey. $\rm\zeta$ is a combination of the TiO5, CaH2, and CaH3 spectral indices (see definition in Equation \ref{equa:zeta}) and has been shown to be correlated with the metallicity. The black curve is an iso-$\rm\zeta$ line with $\rm\zeta=1$, providing an estimate of the TiO-to-CaH ratio in an object of a solar metallicity. Smaller $\rm\zeta$ values correspond to stars of lower metallicities. The index was first introduced in \cite{2007ApJ...669.1235L}, and then recalibrated in \cite{2012AJ....143...67D} and \cite{2013AJ....145..102L}. Table \ref{tab:coe} compares these calibrations.}\label{fig:zeta}
\end{figure}

We measure the spectral indices defined in Section \ref{subsec:basic} and Table 1 on all spectra of the ``labeled'' subdwarf sample and on all spectra of the M-dwarf comparison sample. The M-dwarf results are used to build the TiO5$-$[CaH2+CaH3] diagram shown in Figure \ref{fig:zeta}. The barycentric line across the data cloud directly provides the recalibration of the $\rm\zeta_{TiO/CaH}=1$ curve for average solar metallicity objects. We get:

\begin{equation} \label{equa:new_cah_tio}
\begin{split}
\rm [TiO5]_{Z_\odot}\ &=\ -0.2849\\
&\rm +0.312\times (CaH2\ +\ CaH3)\\
&\rm +0.3863\times (CaH2\ +\ CaH3)^2\\
&\rm -0.1069\times (CaH2\ +\ CaH3)^3
\end{split}
\end{equation}

\begin{deluxetable*}{ccccc}[!ht]
\tablecaption{Coefficients for different calibrations of $\rm [TiO5]_{Z_\odot}$ (See Equation \ref{equa:cah23-tio5} and \ref{equa:new_cah_tio}) \label{tab:coe}}
\tablecolumns{5}
\tablewidth{0pt}
\tablehead{
\colhead{Coefficients} &
\colhead{\cite{2007ApJ...669.1235L}} &
\colhead{\cite{2012AJ....143...67D}} &
\colhead{\cite{2013AJ....145..102L}} &
\colhead{This work} 
}
\startdata
c0 & -0.05   & -0.047  & 0.622  &  -0.2849 \\
c1 & -0.118   & -0.127  & -1.906 & 0.312 \\
c2 & 0.67     & 0.694   & 2.211    & 0.3863\\
c3 & -0.164 & -0.183  & -0.588 & -0.1069 \\
c4 & ...         & -0.005  & ...         & ...\\
\enddata
\end{deluxetable*}

\begin{figure*}
\plotone{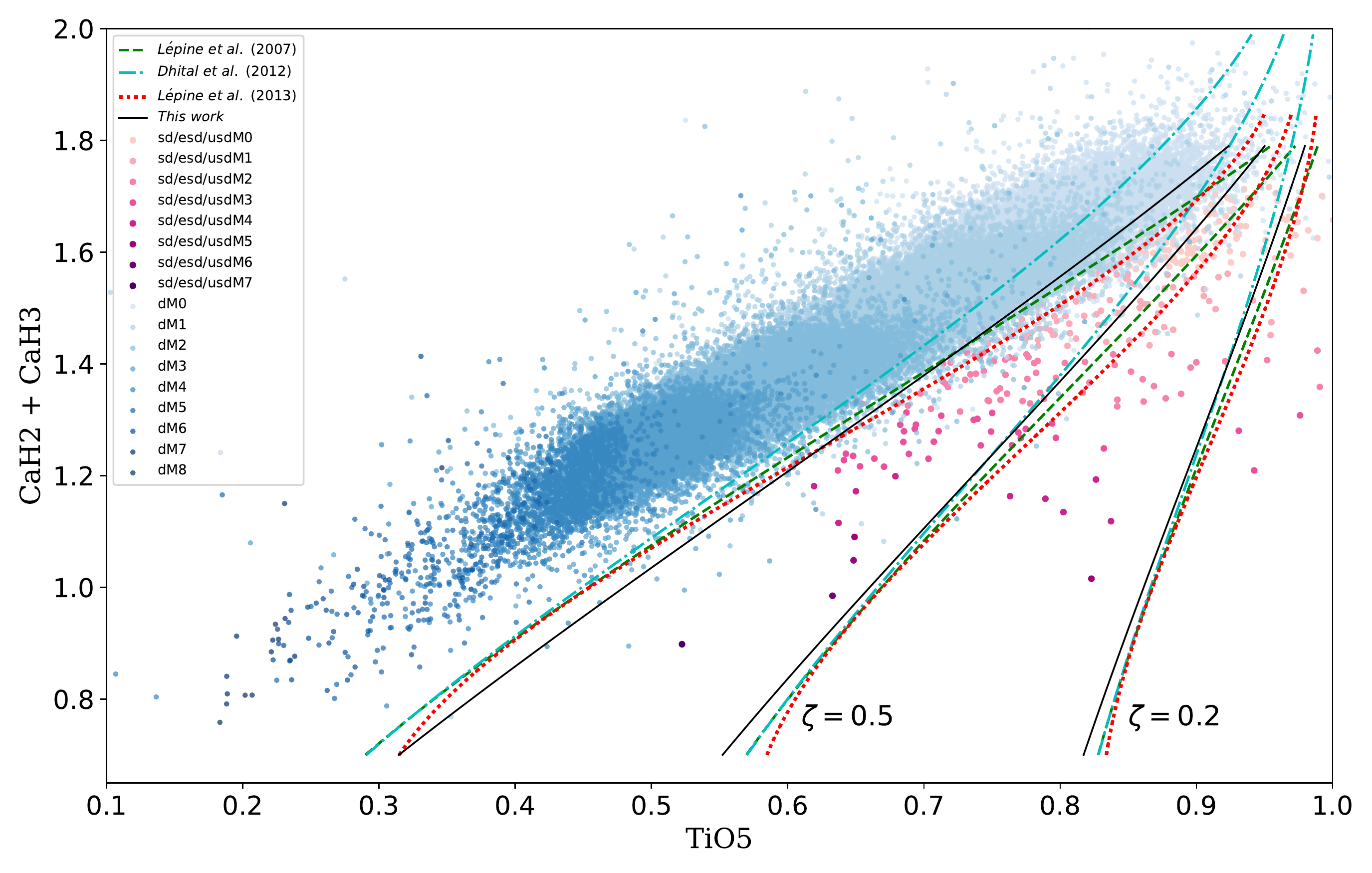}
\caption{New calibration of the $\zeta$ parameter used for subdwarf selection and classification. We plot the ordinary M-dwarf sample (83,213 objects) as blue dots and the ``labeled'' subdwarf sample (433 objects) as pink dots (see Section \ref{subsec:first_sample}) in a [TiO5, CaH2+CaH3] index diagram. The final condition we adopt to classify a target as a ``subdwarf'' is $\zeta<0.75$ (plotted as the solid black curve), because it can exactly cover most of our ``labeled'' subdwarf sample and excludes more than 98\% of the M dwarfs. The iso-$\zeta$ contours from the earlier calibrations of \cite{2007ApJ...669.1235L}, \cite{2012AJ....143...67D} and \cite{2013AJ....145..102L} are also shown for comparison as dashed, dashed-dotted, and dotted lines, respectively. The iso-$\zeta$ curves used to separate subdwarfs from dwarfs in the three former systems correspond to $\zeta<0.825$, as shown in the figure (all different from each other due to the sample selection bias). We formally adopt $\zeta<0.5$ and $\zeta<0.2$ as separators between sdM, esdM, and usdM, following the former systems.}\label{fig:separators}
\end{figure*}

Based on this updated calibration of $\zeta$, we find that the optimal separator curve used to select subdwarfs from dwarfs in the [TiO5, [CaH2+CaH3]] space can be defined as $\zeta<0.75$, as shown in Figure \ref{fig:separators}: it exactly separates almost all our subdwarfs belonging to the ``labeled'' subsample. Furthermore, $\zeta<0.75$ excludes a sufficient fraction, in a statistical sense, of ordinary dwarfs (98\%). This $\zeta$ value is smaller than the canonical 0.825 recommended by L07, \cite{2012AJ....143...67D}, and \cite{2013AJ....145..102L}, which reflects the differences in the instrumental response to the real spectra between LAMOST and other setups. The separators between sdM/esdM/usdM are still adopted as $\zeta<0.5$ and $\zeta<0.2$. Figure \ref{fig:separators} shows the comparison between these several systems.

\subsection{CaH1-Based Dwarf/Subdwarf Separators} \label{subsec:new_cah1}

\begin{figure*}
\center
\includegraphics[width=176mm]{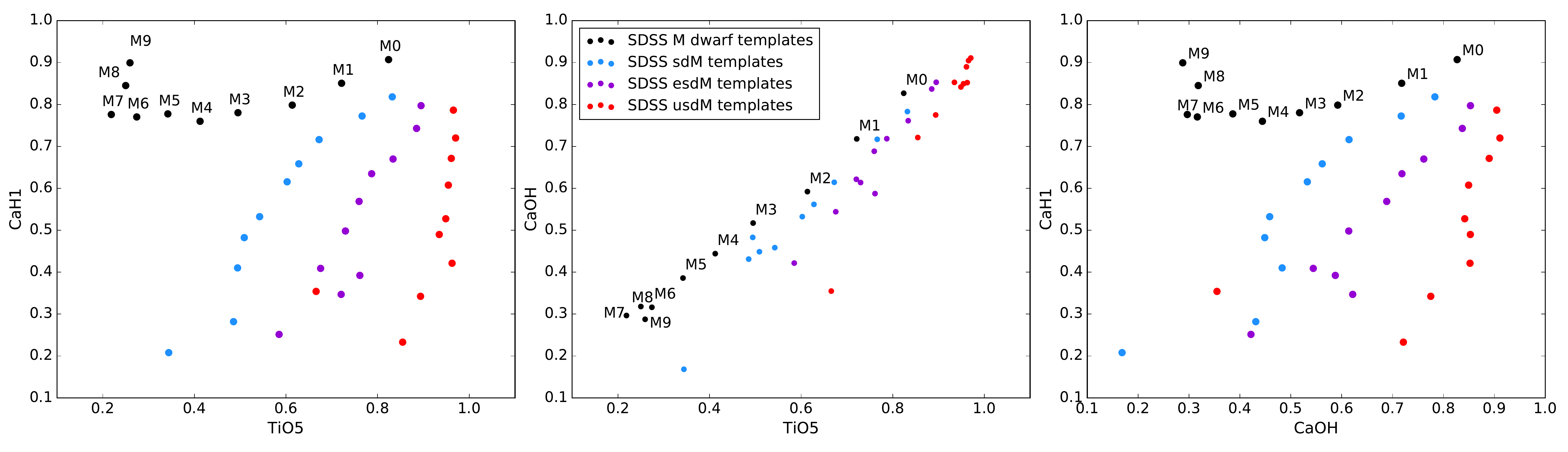}
\caption{Three panels compare the M-dwarf template spectra built by \cite{2007AJ....133..531B} from the Sloan Digital Sky Survey (SDSS) with M-subdwarf template spectra (sdM/esdM/usdM) built by \cite{2014ApJ...794..145S}, also from SDSS, on the [TiO5, CaH1], [TiO5, CaOH], and [CaOH, CaH1] index diagrams, respectively. The left panel shows that M subdwarfs are clearly separated from dwarfs by the CaH1-TiO5 index couple. The central panel shows that the CaOH index variation is consistent with that of the TiO5 index, leading to similar results between CaH1 vs. TiO5 (left panel) and CaH1 vs. CaOH (right panel) distributions. A CaOH-CaH1 relation can thus be defined on the [CaOH, CaH1] index diagram to separate subdwarfs from dwarfs.}\label{fig:templates}
\end{figure*}

The characteristic of prominent CaH1 absorption in subdwarfs can be clearly seen on Figure \ref{fig:templates}, where we compare the index values of M-dwarf and M-subdwarf templates built by \cite{2007AJ....133..531B} and S14, respectively. The result shows excellent discrimination between dwarfs and subdwarfs across the entire spectral sequence when plotting on the [CaH1, TiO5] diagram. Therefore, we suggest that the CaH1 feature should be taken into consideration to select spectroscopic subdwarfs, especially for early-M-type subdwarfs of moderate metal deficiency in which TiO5 absorption is not very strong and $\zeta$ loses efficiency as a separator.

In addition, as the middle diagram of Figure \ref{fig:templates} shows, the CaOH index and the TiO5 index show similar monotonic trend along the effective temperature sequence. The result is to be expected, because the CaOH band depth, as well as the TiO5 one, heavily relies on the element O. Due to the CaOH feature being much closer to the CaH1 band than the TiO5 feature, the ratio of the CaOH index to the CaH1 index maybe less affected by errors arising from the distortion of continuum affected by various instrumental factors or flux calibration problems \citep{2016ApJS..227...27D}.

Using the visually identified dwarfs and ``labeled'' subdwarfs, we derive two equations to define separator curves on the CaH1 versus TiO5 index diagram and CaH1 versus CaOH index diagram, respectively, yielding:

\begin{equation}{\label{equa:cah1_tio5}}
\rm CaH1\ <\ 0.6522\times \rm TiO5^2 -\ 0.577\times \rm TiO5\ +\ 0.8937
\end{equation}
\begin{equation}{\label{equa:cah1_caoh}}
\begin{split}
\rm CaH1 &\rm < 0.4562\times \rm CaOH^3\ -\ 0.1977\times \rm CaOH^2\\
&\rm -\ 0.01899\times \rm CaOH\ -\ 0.7631
\end{split}
\end{equation}

\begin{figure*}
\center
\includegraphics[width=176mm]{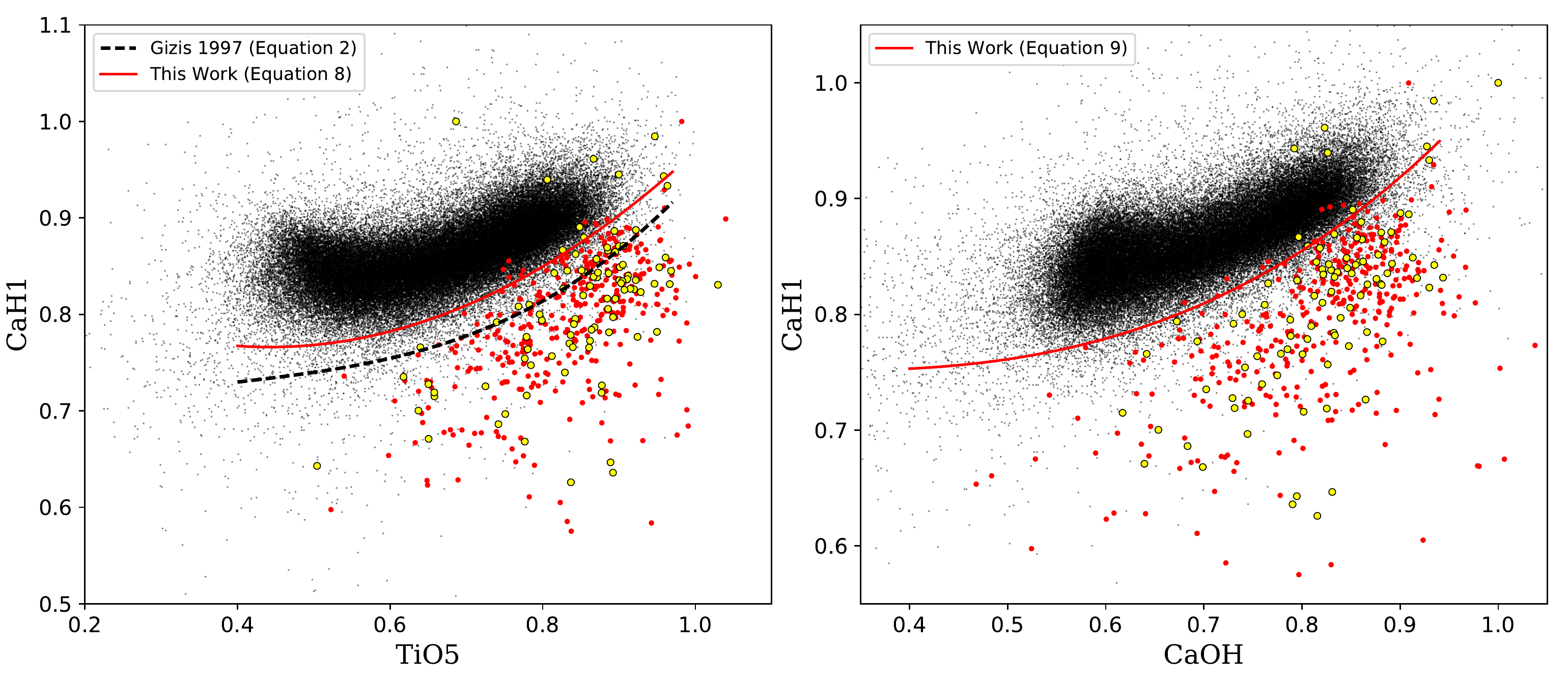}
\caption{To separate M subdwarfs from M dwarfs on the [TiO5, CaH1] and [CaOH, CaH1] index diagrams, we plot the visually identified M dwarfs as black dots, the ``labeled'' subdwarf sample as red dots (see Section \ref{subsec:first_sample}), and the subdwarfs identified by \cite{2016RAA....16..107B} as yellow dots. We define two separator lines (i.e., Equations \ref{equa:cah1_tio5} and \ref{equa:cah1_caoh}) as red solid curves. The dashed black curve on the left panel is the dwarf–subdwarf separator relation originally proposed by \cite{1997AJ....113..806G}, which does not appear suitable for LAMOST data.} \label{fig:cah1_equas}
\end{figure*}

Equation (\ref{equa:cah1_tio5}) is shown in the left panel of Figure \ref{fig:cah1_equas} with a comparison with the original one from G97, i.e., Equation (\ref{equa:g97}). Equation (\ref{equa:cah1_caoh}) is shown in the right panel of Figure\ref{fig:cah1_equas}.

\subsection{Spectral Subtyping} \label{subsec:spt}
The determination of spectral subtypes (SpT) for cool subdwarfs can be made by several methods, such as using the depth of the CaH molecular bands (G97; L07) or the shape of the relatively opacity-free region within 8200-9000 $\rm{\AA}$ \citep{2008AJ....136..840J}. In this work, we follow the definition in L07 in which the ``metallicity'' subclass assignment is dependent on the combined value of the CaH2 and CaH3 indices, given as:

\begin{equation}{\label{equa:spt}}
\rm Sp = 1.4\times (CaH2+CaH3)^2 -10\times(CaH2+CaH3)+12.4
\end{equation}
It will thus be easy to compare our sample with previous works on the spectral sequence. The derived value of SpT is rounded up to the nearest integer, giving the spectral subtype value from 0 for M0 to 7 for M7. 

\subsection{The LAMOST DR4 Subdwarf Final Sample} \label{subsec:final_sample}

To build the final sample, we search by automatic processing LAMOST DR4 using the constraints detailed above. A spectrum is adopted as a candidate if it meets the following three conditions:\\
\\
(1) It has an available RV measurement.\\
(2) $\rm\zeta<0.75$;\\
(3) It satisfies Equation (\ref{equa:cah1_tio5}) or (\ref{equa:cah1_caoh}).\\
\\
Ultimately, visual examination of the spectrum is utilized to confirm (or not) the nature of the candidate.

To avoid missing targets as much as possible, we search for candidate subdwarfs among late-K-type stars, M-type stars, QSOs, and unknowns (as classified by LAMOST 1D automated pipeline), including 1271,426 spectra in total. Adding QSOs to the sample to be explored was decided because of strong molecular absorption by TiO (and VO in the coolest objects) gives a very choppy appearance to the spectra of late-type dwarf stars, quite similar to continuum breaks exhibited by QSOs in specific redshift ranges. As a matter of fact, cool M-type stars and QSO are often contaminating each other in surveys dedicated to each class \citep{1997AJ....113.1421K}.

Our final subdwarf sample is constructed via visual inspection of each candidate spectrum. It consists of 2791 objects and covers the spectral sequence from M0 to M7, in which 291 objects had been investigated by previous researches. Figure \ref{fig:spt} summarizes the distribution of their spectral subtypes. Most objects are earlier than M4, a result not unexpected since the metal deficiency has a tendency to decrease atmospheric opacity and to move the spectral types toward higher effective temperatures, on one hand. On the other hand, instrumental selection by the red channel spectrograph performance of LAMOST also plays against detection of very cool faint objects. According to our newly recalibrated $\rm\zeta$ parameter, the present sample can also be divided into the sdM/esdM/usdM classes with $\rm\zeta<$0.75, $\rm\zeta<$0.5 and $\rm\zeta<$0.2, respectively. There are 2386 sdMs, 295 esdMs, and 110 usdMs in total.

\subsection{The LAMOST DR4 Subdwarf Catalog and External Data} \label{subsec:catalog}

We provide an M-subdwarf catalog containing 2791 objects, the spectra of which can be accessed from the LAMOST Data Release web portall\footnote{\url{http://dr4.lamost.org/}} and the catalog can be obtained online as well as by contacting the corresponding author. We give the values and errors of each spectral index (CaOH, CaH1, CaH2, CaH3, and TiO5), compute the $\zeta$ value, and provide an H$\rm\alpha$ activity indicator for each target. The radial velocities are measured with the method described in Section \ref{subsec:rv}. The spectral subtypes for subdwarfs are computed based on Equation (\ref{equa:spt}) \citep{2007ApJ...669.1235L}.

Additional data from external sources are compiled in this catalog: photometric magnitudes measured in various bandpasses; five optical and near-infrared bands \citep[g, r, i, z, y;][]{2012ApJ...750...99T} from the Pan-STARRS1 (PS1) 3$\pi$ survey \citep{2010SPIE.7733E..0EK}, which is a systematic imaging survey of 3/4 of the sky north of -30$^\circ$; the near-infrared J (1.25 $\mu$m), H (1.65 $\mu$m), and Ks (2.16 $\mu$m) bands from the Two Micron All Sky Survey (2MASS); and W1, W2, W3, and W4 bands centered at wavelengths of 3.4, 4.6, 12, and 22 $\mu$m, respectively, are provided by the AllWISE Data Release \citep{cutri}, which has produced a new Source Catalog and Image Atlas with enhanced sensitivity and accuracy compared with earlier Wide-field Infrared Survey Explorer \citep[WISE;][]{2010AJ....140.1868W} data releases \citep{2014ApJ...783..122K, 2016ApJS..224...36K}. G, BP, and RP magnitudes in the new Gaia photometric system \citep{2018A&A...616A...1G} are also provided.

The astrometric parameters from Gaia DR2 are also provided: proper motions and parallaxes \citep{2018A&A...616A...1G}. We also give estimated distances recorded in the catalog from \cite{2018AJ....156...58B}. An important point must be underlined: the distances provided by \cite{2018AJ....156...58B} are probabilistic estimates based on an elaborate statistical processing of Gaia DR2 data, using specific prior modeling (in particular, stellar density distribution along Galactic lines of sight) to avoid physical errors resulting from ``blind'' simple inversions of measured parallaxes, as discussed in e.g., \cite{2018A&A...616A...9L}.

Table \ref{tab:sample1} gives the description of each column of the catalog. A partial catalog extract is also shown in Table \ref{tab:example}. The complete catalog is available in CSV format online\footnote{\url{http://paperdata.china-vo.org/szhang/DR4_Subdwarfs.csv}}.

\subsection{A By-product: Active Objects Detection from the H$\rm\alpha$ Emission Line} \label{subsec:activity}

Chromospheric activity is common among ordinary M dwarfs and has been largely studied by several authors, in particular, from the large spectroscopic database of the SDSS \citep[see e.g.,][]{2004AJ....128..426W,2011AJ....141...97W} and it has also been found in many M subdwarfs \citep{2007AJ....133..531B, 2014ApJ...794..145S}. At the spectral resolution of survey instruments like the SDSS combination or LAMOST, the brightest and most easily detectable emission line is H$\rm\alpha$, accompanied in the most active objects by Ca II emission (both in H and K lines and in the far-red triplet), and sometimes by higher Balmer series lines. According to the criteria for designating a star as active as defined by \cite{2011AJ....141...97W}, we measure the equivalent widths of an H$\rm\alpha$ emission line for each object and remove the unreliable ones via visual inspection. Finally, a total of 141 active subdwarfs with prominent H$\rm\alpha$ emission lines are found, including 120 sdMs, 18 esdMs, and 3 usdMs, amounting to more than 5\% of the total sample. The H$\rm\alpha$ activity flag for each object is listed in the catalog. The flag is set as 1 for the active ones, 0 for the ones without emission features, and -9999 for the ones with S/N$<$3 in the r or i band.

\begin{figure}
\center
\includegraphics[width=84mm]{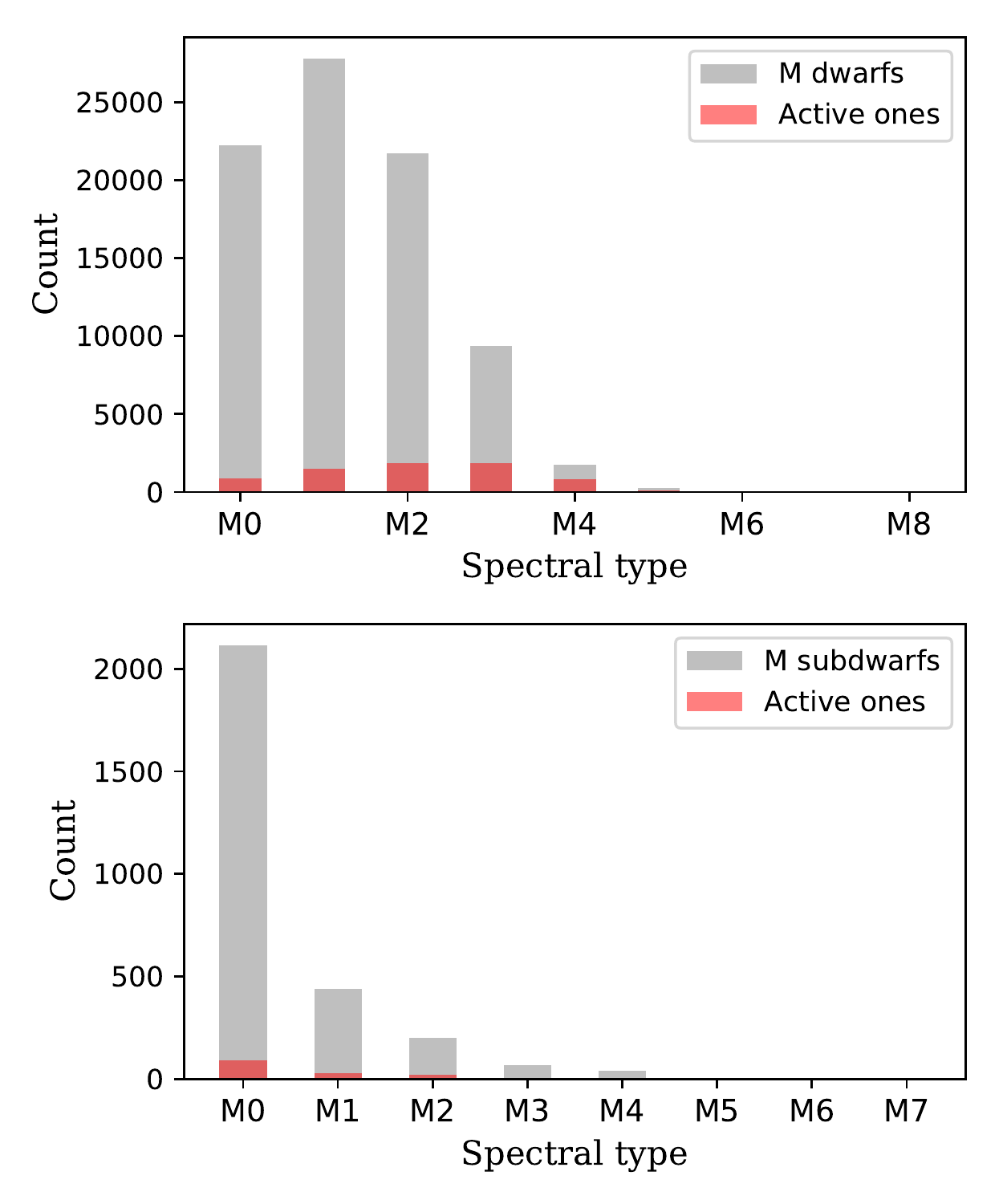}
\caption{SpT distribution of visually identified M dwarfs (83,213 in total, top panel) compared with SpT distribution of subdwarfs (bottom panel). The magnetic activity fraction for each spectral subtype, detected by H$\rm\alpha$ emission following the precepts of \cite{2011AJ....141...97W}, is shown as red bars. We found 7075 active M dwarfs and 141 M subdwarfs. The figure shows that the magnetic activity fraction is a function of the spectral type for M dwarfs, which is consistent with former studies \citep{2011AJ....141...97W}. For subdwarfs, the active subsample remains too small to draw a reliable conclusion.} \label{fig:spt}
\end{figure}

As Figure \ref{fig:spt} shows, the activity fraction of M dwarfs is increasing in later types, a result consistent with former studies \citep[see e.g.,][]{2011AJ....141...97W}. Unfortunately, due to the observational selection against subdwarfs cooler than M3-M4 in our sample, it is difficult to derive a reliable estimate of the fraction of active subdwarfs. Previous studies have shown that among M subdwarfs, chromospheric activity is significantly less frequent than among ordinary M dwarfs, but the spectral type at which the maximum rate of activity observed does not appear to differ. This reduction is claimed to be an argument to confirm that the chromospheric activity is a proxy for the age of the subdwarfs, whose bulk of observationally selected population that are associated with the Galactic halo and thick disk should be of substantially older formation than that of the disk dwarfs.

\section{Summary and Conclusion} \label{sec:conclude}

In order to search for spectroscopic cool subdwarfs from DR4 of the LAMOST regular survey, we review the history of current classification system for M subdwarfs developed by \cite{2007ApJ...669.1235L} and suggested to make some modifications based on the analysis of subdwarf spectra and templates from \cite{2014ApJ...794..145S}. We suggest that the [CaH1, TiO5] index relation originally proposed by \cite{1997AJ....113..806G} should be kept as a necessary condition for safe spectroscopic identification of subdwarfs, in combination with the classical [TiO5, CaH2 +CaH3] index relation.

To adapt the molecular index relations to LAMOST data as separators between dwarfs and subdwarfs, we built a sample of more than 80,000 M-dwarf spectra from LAMOST regular survey verified by means of the manual ``eyecheck'' mode of the Hammer spectral typing utility. Meanwhile, a `labeled'' subdwarf sample consists of 325 targets from this procedure and 108 subdwarfs from LAMOST DR2 in \cite{2016RAA....16..107B}. Using the dwarf sample and ``labeled'' subdwarf sample, we revise the separator relation on the [TiO5 versus CaH1] diagram and define a new separator on the [CaOH versus CaH1] diagram. Further, the $\rm\zeta$ index is recalibrated from the [TiO5, CaH2+CaH3] relation obeyed by LAMOST M dwarfs. The best new separator $\rm\zeta$ value between dwarfs and subdwarfs is found to be 0.75. The spectral subtypes of subdwarfs are determined by the compound CaH2+CaH3 index.

Using these revised schemes as filters and visual inspection as the final verification tool, we identify 2791 M subdwarfs from the entire LAMOST DR4, out of more than a million of spectra submitted to the screening. In this sample, 141 have detectable H$\rm\alpha$ emissions.

For a more convenient use of this sample, the catalog also provides the astrometric information from Gaia DR2, the distance estimated by \cite{2018AJ....156...58B}, and a variety of magnitudes, including photometric magnitudes from Pan-STARRS1, 2MASS, AllWISE, and Gaia. The catalog is provided online, and the spectra can be accessed from the LAMOST Data Release web portal.

The spectroscopic selection and classification systems were originally built from objects with ``old population'' kinematic properties. However, the reciprocal statement, i.e., whether the large ``subdwarfs'' samples selected and classified using spectral indices all exhibit these kinematic properties is still an open question to explore. Moreover, the subdwarfs appearing in directions corresponding to the Galactic thin disk are especially interesting, because subdwarfs are believed to evolve extremely slowly and most of those already identified are residents of the halo and the thick disk. Is this a pure projection effect or have these objects migrated or been trapped in the thin disk? Therefore, the correlations between their metallicity, true spatial location, and kinematics might reveal crucial information on the formation and evolution history of our the Milky Way. Our M-subdwarf sample contains a large number of objects located at low Galactic latitudes especially in the Galactic anti-center direction, allowing exploration of differential kinematic properties with respect to previously published halo- and thick disk-dominated samples (S. Zhang 2019, in preparation). In addition, this sample also supplies candidates for high-resolution observations in the future.

We thank Andrew A. West for helpful discussion and for the supply of the latest version of catalog in \cite{2011AJ....141...97W}. We also gratefully appreciate the penetrating comments from the referee that drastically improved the manuscript. 

This work was funded by the National Basic Research Program of China (973 Program, 2014CB845700) and the National Natural Science Foundation of China (Grant Nos. 11390371, 11703053). Guoshoujing Telescope (the Large Sky Area Multi-Object Fiber Spectroscopic Telescope LAMOST) is a National Major Scientific Project built by the Chinese Academy of Sciences. Funding for the project has been provided by the National Development and Reform Commission. LAMOST is operated and managed by the National Astronomical Observatories, Chinese Academy of Sciences.


\software{ Hammer \citep{2007AJ....134.2398C},
		Astropy \citep{2013A&A...558A..33A},
		Matplotlib \citep{2005ASPC..347...91B},
		Pandas (http://pandas.pydata.org),
		Numpy (http://www.numpy.org)
        } 
        
\appendix

\begin{deluxetable}{llll}[ht!]
\tablecaption{Data description of the subdwarf catalog \label{tab:sample1}}
\tablewidth{0pt}
\tablenum{4}
\tablehead{
\colhead{Column Name} & \colhead{Unit\qquad \quad  } & \colhead{Description\qquad \qquad \qquad \qquad \qquad \qquad} }
\startdata
designation & - & Target Designation \\
fitsname & - & Spectrum name \\
mjd & - & Day  \\ 
planid & - & Plan Name \\
spid & - & Spectrograph ID \\
fiberid & - & Fiber ID \\
ra & degrees & Right Ascension \\
dec & degrees & Declination \\
\enddata
\end{deluxetable}
 
\begin{deluxetable}{llll}[ht!]
\scriptsize
\tablecaption{Data description of the subdwarf catalog \label{tab:sample2}}
\tablewidth{0pt}
\tablenum{4}
\tablehead{
\colhead{Column Name\qquad\qquad\qquad} & \colhead{Unit\qquad \qquad\qquad\qquad} & \colhead{Description\qquad \qquad \qquad \qquad \qquad \qquad\qquad\qquad\qquad\qquad\qquad\qquad\qquad\qquad\qquad\qquad} }
\startdata
l & degrees & Galactic longitude\\
b & degrees & Galactic latitude \\
snrr & - & Average spectrum signal to noise Ratio in r band  \\
snri & - & Average spectrum signal to noise Ratio in i band \\
rv & km s$^{-1}$ & Heliocentric radial velocity  \\
rv\_err & km s$^{-1}$ & Standard error of radial velocity  \\
CaOH & - & Spectral index of CaOH band \\
CaOH\_err & - & The error of spectral index of CaOH band \\
CaH1 & - & Spectral index of CaH1 band  \\
CaH1\_err & - & The error of spectral index of CaH1 band \\
CaH2 & - & Spectral index of CaH2 band  \\
CaH2\_err & - & The error of spectral index of CaH2 band \\
CaH3 & - & Spectral index of CaH3 band  \\
CaH3\_err & - & The error of spectral index of CaH3 band \\
TiO5 & - & Spectral index of TiO5 band  \\
TiO5\_err & - & The error of spectral index of TiO5 band \\
Zeta & - & The value of $\zeta$ parameter  \\
spt & - & Spectral subtype \\
activity & - & H$\alpha$ emission tag  \\
flag\_caution & - & A flag for spectrum quality from visual inspection  \\
parallax & mas & Gaia DR2 Parallax  \\
parallax\_error & mas & Standard error of Gaia DR2 parallax \\
pmra & mas yr$^{-1}$ & Gaia DR2 proper motion in right ascension direction\\
pmra\_error & mas yr$^{-1}$  & Standard error of Gaia DR2 proper motion in right ascension direction\\
pmdec & mas yr$^{-1}$ & Gaia DR2 proper motion in declination direction\\
pmdec\_error & mas yr$^{-1}$ & Standard error of Gaia DR2 proper motion in declination direction\\
phot\_g\_mean\_mag & mag & Gaia DR2 G-band mean magnitude \\
phot\_bp\_mean\_mag & mag & Gaia DR2 integrated BP mean magnitude \\
phot\_rp\_mean\_mag & mag & Gaia DR2 integrated RP mean magnitude\\
bp\_rp & mag & Gaia DR2 BP-RP colour.  \\
rest & pc & Estimated distance from \citet{2018AJ....156...58B}, based on  Gaia DR2 \\
b\_rest & pc & Lower bound on the confidence interval of the estimated distance \\
B\_rest & pc & Upper bound on the confidence interval of the estimated distance\\
rlen & pc & Length scale used in the prior for the distance estimation\\
gmag & AB magnitudes & Pan-STARRS1 mean PSF magnitude from g filter detections \\
e\_gmag & AB magnitudes & Error in Pan-STARRS1 g magnitude detections\\
rmag & AB magnitudes &  Pan-STARRS1 mean PSF magnitude from r filter detections \\
e\_rmag & AB magnitudes &Error in Pan-STARRS1 r magnitude detections\\
\enddata
\end{deluxetable}

\begin{deluxetable}{llll}[ht!]
\tablecaption{Data description of the subdwarf catalog \label{tab:sample3}}
\tablewidth{0pt}
\tablenum{4}
\tablehead{
\colhead{Column Name\qquad} & \colhead{Unit\qquad \quad  } & \colhead{Description\qquad \qquad \qquad \qquad \qquad \qquad} }
\startdata
imag & AB magnitudes &  Pan-STARRS1 mean PSF magnitude from i filter detections\\
e\_imag & AB magnitudes & Error in Pan-STARRS1 i magnitude detections\\
zmag & AB magnitudes &  Pan-STARRS1 mean PSF magnitude from z filter detections \\
e\_zmag & AB magnitudes & Error in Pan-STARRS1 z magnitude detections\\
ymag & AB magnitudes &  Pan-STARRS1 mean PSF magnitude from y filter detections \\
e\_ymag & AB magnitudes & Error in Pan-STARRS1 y magnitude detections\\
Jmag & mag & 2MASS default J-band magnitude  \\
Hmag & mag & 2MASS default H-band magnitude\\
Kmag & mag & 2MASS default Ks-band magnitude\\
e\_Jmag & mag & Photometric uncertainty for the 2MASS J-band magnitude \\
e\_Hmag & mag &Photometric uncertainty for the 2MASS H-band magnitude \\
e\_Kmag & mag &Photometric uncertainty for the 2MASS Ks-band magnitude\\
W1mag & mag &  AllWISE W1 standard aperture magnitude\\
W2mag & mag & AllWISE W2 standard aperture magnitude\\
W3mag & mag & AllWISE W3 standard aperture magnitude\\
W4mag & mag & AllWISE W4 standard aperture magnitude\\
e\_W1mag & mag & Uncertainty in the AllWISE W1 standard aperture magnitude\\
e\_W2mag & mag & Uncertainty in the AllWISE W2 standard aperture magnitude\\
e\_W3mag & mag & Uncertainty in the AllWISE W3 standard aperture magnitude\\
e\_W4mag & mag & Uncertainty in the AllWISE W4 standard aperture magnitude\\
\enddata
\tablecomments{The complete LAMOST DR4 M subdwarf catalog can be downloaded from the on-line journal. Columns 2 to 12 record the information provided by LAMOST data base and the spectrum FITS files headers. Radial velocities, spectral indices and subtypes are measured in this work. The activity flag is used to indicate the magnetic activity of an object, which is measured by H$\alpha$ emission line. A flag\_caution is provided based on the visual inspection procedure: the objects with low quality spectra or inaccurate flux calibration are set as ``*'' to warn that the data on these objects should be used carefully. }
\end{deluxetable}

\begin{longrotatetable}
\begin{deluxetable*}{rlllllllllllll}
\tablecaption{A partial extract of the LAMOST DR4 M subdwarf catalog\label{tab:example}}
\tablewidth{700pt}
\tabletypesize{\scriptsize}
\tablenum{5}
\tablehead{
\colhead{designation} & \colhead{fitsname} & \colhead{mjd} & \colhead{planid} & \colhead{spid} & \colhead{fiberid} & \colhead{ra} & \colhead{dec} & \colhead{l} & \colhead{b} \\
\colhead{}& \colhead{snrr}& \colhead{snri}&\colhead{rv} & \colhead{rv$\_$err} & \colhead{CaOH} & \colhead{CaOH$\_$err} & \colhead{CaH1} & \colhead{CaH1$\_$err} & \colhead{CaH2}\\
\colhead{} & \colhead{CaH2$\_$err} & \colhead{CaH3}&  \colhead{CaH3$\_$err} & \colhead{TiO5} & \colhead{TiO5$\_$err}  &  \colhead{Zeta}  & \colhead{spt} & \colhead{activity} & \colhead{flag$\_$caution} & \colhead{}& \colhead{}
} 
\startdata
J002531.37+355526.4 & spec-56952-M31005N35M1$\_$sp09-206 & 56952 & M31005N35M1 & 9 & 206 & 6.380737 & 35.924017 & 117.0639812 & -26.65228601 \\
 &14.9 & 32.2 & 37 & 36 & 0.729 & 0.029 & 0.833 & 0.022 & 0.699\\
 & 0.015 & 0.875 & 0.018 & 0.828 & 0.02 & 0.677 & 0 & 0 &   \\
J003921.57+345754.2 & spec-55910-M31$\_$007N34$\_$B2$\_$sp10-129 & 55910 & M31$\_$007N34$\_$B2 & 10 & 129 & 9.839904 & 34.965069 & 120.1336293 & -27.84104438\\
 & 16 & 40.6 & 52 & 8 & 0.777 & 0.023 & 0.844 & 0.02 & 0.693\\
 & 0.012 & 0.851 & 0.014 & 0.798 & 0.015 & 0.733 & 0 & 0 &  \\
J004613.83+335010.3 & spec-55910-M31$\_$007N34$\_$B2$\_$sp05-222 & 55910 & M31$\_$007N34$\_$B2 & 5 & 222 & 11.55766 & 33.836202 & 121.6953138 & -29.02304224\\
 & 11.8 & 32.6 & 48 & 28 & 0.91 & 0.038 & 0.774 & 0.026 & 0.567\\
 & 0.012 & 0.757 & 0.014 & 0.842 & 0.018 & 0.357 & 1 & 0 &  \\
J005303.21+271507.1 & spec-56551-VB010N27V2$\_$sp06-142 & 56551 & VB010N27V2 & 6 & 142 & 13.263398 & 27.251989 & 123.3736522 & -35.61837284\\
 & 9.9 & 24.2 & 34 & 12 & 0.701 & 0.04 & 0.776 & 0.034 & 0.474\\
 & 0.014 & 0.706 & 0.018 & 0.695 & 0.022 & 0.55 & 2 & 0 &  \\
J011701.85+335911.3 & spec-57373-M31019N31B1$\_$sp11-217 & 57373 & M31019N31B1 & 11 & 217 & 19.25772 & 33.986477 & 128.9722096 & -28.58492645\\
 & 90.7 & 147.6 & 29 & 13 & 0.787 & 0.004 & 0.857 & 0.004 & 0.719\\
 & 0.003 & 0.859 & 0.003 & 0.816 & 0.004 & 0.735 & 0 & 0 &  \\
J012534.63-041817.6 & spec-56591-EG012606S021203F01$\_$sp01-034 & 56591 & EG012606S021203F01 & 1 & 34 & 21.394302 & -4.304911 & 144.0657935 & -65.7660619\\
 & 10.1 & 17.4 & 90 & 11 & 0.802 & 0.042 & 0.78 & 0.036 & 0.674\\
 & 0.03 & 0.816 & 0.034 & 0.845 & 0.038 & 0.492 & 0 & 0 &  \\
J012615.42+352030.6 & spec-56657-M31020N36M1$\_$sp08-104 & 56657 & M31020N36M1 & 8 & 104 & 21.564258 & 35.341851 & 130.8954916 & -26.99090458\\
 & 12.5 & 28.4 & -105 & 11 & 0.812 & 0.034 & 0.773 & 0.024 & 0.655\\
 & 0.016 & 0.836 & 0.018 & 0.843 & 0.02 & 0.498 & 0 & 0 &  \\
J012706.90-032656.1 & spec-56233-EG012606S021203B03$\_$sp05-009 & 56233 & EG012606S021203B03 & 5 & 9 & 21.778777 & -3.44894 & 144.2726261 & -64.83205927\\
 & 7.3 & 28.1 & 27 & 11 & 0.641 & 0.046 & 0.815 & 0.04 & 0.596\\
 & 0.019 & 0.81 & 0.023 & 0.72 & 0.026 & 0.738 & 1 & 0 & * \\
J013745.91+321343.0 & spec-56650-M31021N31M1$\_$sp13-050 & 56650 & M31021N31M1 & 13 & 50 & 24.441311 & 32.228621 & 134.1987802 & -29.62755946\\
 & 12.2 & 23.2 & -47 & 14 & 0.702 & 0.038 & 0.842 & 0.026 & 0.678\\
 & 0.023 & 0.824 & 0.027 & 0.785 & 0.028 & 0.702 & 0 & 0 & * \\
J022329.68+034930.6 & spec-56299-EG023131N032619F04$\_$sp14-192 & 56299 & EG023131N032619F04 & 14 & 192 & 35.873708 & 3.825187 & 162.1417055 & -51.89742013 \\
 & 16.1 & 35.9 & -29 & 22 & 0.815 & 0.028 & 0.873 & 0.02 & 0.756 \\
 & 0.014 & 0.891 & 0.015 & 0.872 & 0.018 & 0.641 & 0 & 0 &  \\
\enddata
\tablecomments{An extract of the 28 first columns of catalog records on 10 objects.  The full catalog can be accessed on-line.}
\end{deluxetable*}
\end{longrotatetable}


\begin{thebibliography}{}
\bibitem[Abazajian et al.(2009)]{2009ApJS..182..543A} Abazajian, K.~N., Adelman-McCarthy, J.~K., Ag{\"u}eros, M.~A., et al.\ 2009, \apjs, 182, 543-558 
\bibitem[Anguiano et al.(2018)]{2018arXiv180707625A} Anguiano, B., Majewski, S.~R., Allende-Prieto, C., et al.\ 2018, arXiv:1807.07625 
\bibitem[Ake \& Greenstein(1980)]{1980ApJ...240..859A} Ake, T.~B., Greenstein, J.~L.\ 1980, \apj, 240,859
\bibitem[Astropy Collaboration et al.(2013)]{2013A&A...558A..33A} Astropy Collaboration, Robitaille, T.~P., Tollerud, E.~J., et al.\ 2013, \aap, 558, A33 
\bibitem[Bai et al.(2016)]{2016RAA....16..107B} Bai, Y., Luo, A.-L., Comte, G., et al.\ 2016, Research in Astronomy and Astrophysics, 16, 107
\bibitem[Bailer-Jones et al.(2018)]{2018AJ....156...58B} Bailer-Jones, C.~A.~L., Rybizki, J., Fouesneau, M., Mantelet, G., \& Andrae, R.\ 2018, \aj, 156, 58
\bibitem[Barrett et al.(2005)]{2005ASPC..347...91B} Barrett, P., Hunter, J., Miller, J.~T., Hsu, J.-C., \& Greenfield, P.\ 2005, Astronomical Data Analysis Software and Systems XIV, 347, 91 
\bibitem[Bessell(1982)]{1982PASAu...4..417B} Bessell, M.~S.\ 1982, \pasa, 4,417
\bibitem[Bochanski et al.(2007)]{2007AJ....133..531B} Bochanski, J.~J., West, A.~A., Hawley, S.~L., et al.\ 2007, \aj, 133,531
\bibitem[Bochanski et al.(2010)]{2010AJ....139.2679B} Bochanski, J.~J., Hawley, S.~L.,  Covey, K.~R.\ 2010, \aj, 139,2679
\bibitem[Bochanski et al.(2013)]{2013AJ....145...40B} Bochanski, J.~J., Savcheva, A., West, A.~A.\ 2013, \aj, 145,40 (B13)
\bibitem[Burgasser \& Kirkpatrick(2006)]{2006ApJ...645.1485B} Burgasser, A.~J., \& Kirkpatrick, J.~D.\ 2006, \apj, 645, 1485 
\bibitem[Chabrier(2003)]{2003ApJ...586L.133C} Chabrier, G.\ 2003, \apj, 586,133
\bibitem[Covey et al.(2007)]{2007AJ....134.2398C} Covey, K.~R., Ivezic, Z., Schlegel, D., et al.\ 2007, \aj, 134,2398
\bibitem[Cropper et al.(2018)]{2018A&A...616A...5C} Cropper, M., Katz, D., Sartoretti, P., et al.\ 2018, \aap, 616, A5
\bibitem[Cui et al.(2012)]{2012RAA....12.1197C} Cui, X.-Q., Zhao, Y.-H., Chu, Y.-Q., et al.\ 2012, Research in Astronomy and Astrophysics, 12, 1197
\bibitem[Cutri et al.(2014)]{cutri}Cutri, R. M., Wright, E. L., Conrow, T., et al. 2014, http://wise2.ipac.caltech.edu/docs/release/allwise/expsup/
\bibitem[Dhital et al.(2012)]{2012AJ....143...67D} Dhital, S., West, A.~A., Stassun, K.~G., et al.\ 2012, \aj, 143, 67 
\bibitem[Du et al.(2016)]{2016ApJS..227...27D} Du, B., Luo, A.-L., Kong, X., et al.\ 2016, \apjs, 227, 27 
\bibitem[Eggen \& Greenstein(1965)]{1965ApJ...142..925E} Eggen, O.~J. \& Greenstein, J.~L.\ 1965, \apj, 142, 925
\bibitem[Finkbeiner et al.(2016)]{2016ApJ...822...66F} Finkbeiner, D.~P., Schlafly, E.~F., Schlegel, D.~J., et al.\ 2016, \apj, 822, 66 
\bibitem[Fuchs et al.(2009)]{2009AJ....137.4149F} Fuchs, B., Dettbarn, C., Rix, H.-W., et al.\ 2009, \aj, 137, 4149 
\bibitem[Gaia Collaboration et al.(2018)]{2018A&A...616A...1G} Gaia Collaboration, Brown, A.~G.~A., Vallenari, A., et al.\ 2018, \aap, 616, A1
\bibitem[Giclas et al.(1971)]{1971lpms.book.....G} Giclas, H.~L., Burnham, R., \& Thomas, N.~G.\ 1971, Flagstaff, Arizona: Lowell Observatory, 1971,  
\bibitem[Gizis(1997)]{1997AJ....113..806G} Gizis, J.~E.\ 1997, \aj, 113,806 (G97)
\bibitem[Green et al.(2015)]{2015ApJ...810...25G} Green, G.~M., Schlafly, E.~F., Finkbeiner, D.~P., et al.\ 2015, \apj, 810, 25 
\bibitem[Guo et al.(2015)]{2015RAA....15.1182G} Guo, Y.-X., Yi, Z.-P., Luo, A.-L., et al.\ 2015, Research in Astronomy and Astrophysics, 15, 1182 
\bibitem[Hartwick et al.(1984)]{1984ApJ...286..269H} Hartwick, F.~D.~A., Cowley, A.~P., Mould, J.~R.\ 1984, \apj, 286,269
\bibitem[Huang et al.(2015)]{2015MNRAS.449..162H} Huang, Y., Liu, X.-W., Yuan, H.-B., et al.\ 2015, \mnras, 449, 162 
\bibitem[Jao et al.(2008)]{2008AJ....136..840J} Jao, W.-C., Henry, T.~J., Beaulieu, T.~D., \& Subasavage, J.~P.\ 2008, \aj, 136, 840
\bibitem[Jao(2011)]{2011ASPC..448..907J} Jao, W.-C.\ 2011, 16th Cambridge Workshop on Cool Stars, Stellar Systems, and the Sun, 448, 907 
\bibitem[Jao et al.(2016)]{2016AJ....152..153J} Jao, W.~C., Nelan, E.~P., Henry, T.~J.\ 2016, \aj, 152,153
\bibitem[Joy(1947)]{1947ApJ...105...96J} Joy, A.~H.\ 1947, \apj, 105,96
\bibitem[Kafle et al.(2014)]{2014ApJ...794...59K} Kafle, P.~R., Sharma, S., Lewis, G.~F., \& Bland-Hawthorn, J.\ 2014, \apj, 794, 59 
\bibitem[Kaiser et al.(2010)]{2010SPIE.7733E..0EK} Kaiser, N., Burgett, W., Chambers, K., et al.\ 2010, \procspie, 7733, 77330E 
\bibitem[Kirkpatrick et al.(1991)]{1991ApJS...77..417K} Kirkpatrick, J.~D., Henry, T.~J., McCarthy, D.~W.~J.\ 1991, \apjs, 77,417
\bibitem[Kirkpatrick et al.(1997)]{1997AJ....113.1421K} Kirkpatrick, J.~D., Henry, T.~J., Irwin, M.~J.\ 1997, \aj, 113,1421
\bibitem[Kirkpatrick et al.(2014)]{2014ApJ...783..122K} Kirkpatrick, J.~D., Schneider, A., Fajardo-Acosta, S., et al.\ 2014, \apj, 783, 122 
\bibitem[Kirkpatrick et al.(2016)]{2016ApJS..224...36K} Kirkpatrick, J.~D., Kellogg, K., Schneider, A.~C., et al.\ 2016, \apjs, 224, 36 
\bibitem[Kuiper(1939)]{1939ApJ....89..548K} Kuiper, G.~ P.\ 1939, \apj, 89, 548
\bibitem[Laughlin et al.(1997)]{1997ApJ...491L..51L} Laughlin, G., Adams, F.~C.\ 1997, \apj, 491,51
\bibitem[L{\'e}pine et al.(2003a)]{2003AJ....125.1598L} L{\'e}pine, S., Rich, R.~M., Shara, M.~M.\ 2003, \aj, 125, 1598
\bibitem[L{\'e}pine et al.(2003b)]{2003ApJ...585L..69L} L{\'e}pine, S., Shara, M.~M., Rich, R.~M.\ 2003, \apj, 585, 69
\bibitem[L{\'e}pine \& Shara(2005)]{2005AJ....129.1483L} L{\'e}pine, S., \& Shara, M.~M.\ 2005, \aj, 129, 1483 
\bibitem[L{\'e}pine et al.(2007)]{2007ApJ...669.1235L} L{\'e}pine, S., Rich, R.~M., Shara, M.~M.\ 2007, \apj, 669,1235 (L07)
\bibitem[L{\'e}pine \& Scholz(2008)]{2008ApJ...681L..33L} L{\'e}pine, S., \& Scholz, R.-D.\ 2008, \apjl, 681, L33 
\bibitem[L{\'e}pine et al.(2013)]{2013AJ....145..102L} L{\'e}pine, S., Hilton, E.~J., Mann, A.~W., et al.\ 2013, \aj, 145, 102
\bibitem[Lee et al.(2008)]{2008AJ....136.2022L} Lee, Y.~S., Beers, T.~C., Sivarani, T., et al.\ 2008, \aj, 136, 2022 
\bibitem[Liebert et al.(1979)]{1979ApJ...233..226L} Liebert, J., Dahn, C.~C., Gresham, M., Strittmatter, P.~A., et al.\ 1979, \apj, 233,226
\bibitem[Luo et al.(2015)]{2015RAA....15.1095L} Luo, A.-L., Zhao, Y.-H., Zhao, G., et al.\ 2015, Research in Astronomy and Astrophysics, 15, 1095
\bibitem[Luri et al.(2018)]{2018A&A...616A...9L} Luri, X., Brown, A.~G.~A., Sarro, L.~M., et al.\ 2018, \aap, 616, A9
\bibitem[Luyten(1979)]{1979lccs.book.....L} Luyten, W.~J.\ 1979, Minneapolis: University of Minnesota, 1979, 2nd ed.,  
\bibitem[Mould(1976a)]{1976ApJ...207..535M} Mould, J.~R.\ 1976, \apj, 207,535
\bibitem[Mould(1976b)]{1976ApJ...210..402M} Mould, J.~R.\ 1976, \apj, 210,402
\bibitem[Mould \& McElroy(1978)]{1978ApJ...220..935M} Mould, J.~R.,\& McElroy, D.~B.\ 1978, \apj, 220,935
\bibitem[Olmsted(1908)]{1908CMWCI..21....1O} Olmsted, C.~M.\ 1908, Contributions from the Mount Wilson Observatory / Carnegie Institution of Washington, 21, 1 
\bibitem[{\"O}hman(1934)]{1934ApJ....80..171O} {\"O}hman, Y.\ 1934, \apj, 80, 171 
\bibitem[Rajpurohit et al.(2013)]{2013A&A...556A..15R} Rajpurohit, A.~S., Reyl?(C), C., Allard, F., et al.\ 2013, \aap, 556,15
 \bibitem[Rajpurohit et al.(2016)]{2016A&A...596A..33R} Rajpurohit, A.~S., Reyl{\'e}, C., Allard, F., et al.\ 2016, \aap, 596, A33
\bibitem[Reid et al.(1995)]{1995AJ....110.1838R} Reid, I.~N., Hawley, S.~L., Gizis, J.~E.\ 1995, \aj, 110, 1838
\bibitem[Reid et al.(2002)]{2002AJ....124.2721R} Reid, I.~N., Gizis, J.~E., Hawley, S.~L.\ 2002, \aj, 124, 2721
\bibitem[Roeser et al.(2010)]{2010AJ....139.2440R} Roeser, S., Demleitner, M., \& Schilbach, E.\ 2010, \aj, 139, 2440 
\bibitem[Savcheva et al.(2014)]{2014ApJ...794..145S} Savcheva, A.~S., West, A.~A., Bochanski, J.~J., et al.\ 2014, \apj, 794,145 (S14)
\bibitem[Sch{\"o}nrich et al.(2010)]{2010MNRAS.403.1829S} Sch{\"o}nrich, R., Binney, J., \& Dehnen, W.\ 2010, \mnras, 403, 1829 
\bibitem[Skrutskie et al.(2006)]{2006AJ....131.1163S} Skrutskie, M.~F., Cutri, R.~M., Stiening, R., et al.\ 2006, \aj, 131,1163
\bibitem[Tonry \& Davis(1979)]{1979AJ.....84.1511T} Tonry, J., \& Davis, M.\ 1979, \aj, 84, 1511
\bibitem[Tonry et al.(2012)]{2012ApJ...750...99T} Tonry, J.~L., Stubbs, C.~W., Lykke, K.~R., et al.\ 2012, \apj, 750, 99 
\bibitem[West et al.(2004)]{2004AJ....128..426W} West, A.~A., Hawley, S.~L., Walkowicz, L.~M., et al.\ 2004, \aj, 128, 426 
\bibitem[West et al.(2011)]{2011AJ....141...97W} West, A.~A., Morgan, D.~P., Bochanski, J.~J., et al.\ 2011, \aj, 141, 97
\bibitem[Wright et al.(2010)]{2010AJ....140.1868W} Wright, E.~L., Eisenhardt, P.~R.~M., Mainzer, A.~K., et al.\ 2010, \aj, 140, 1868 
\bibitem[Yi et al.(2014)]{2014AJ....147...33Y} Yi, Z., Luo, A., Song, Y., et al.\ 2014, \aj, 147, 33 
\bibitem[Yi et al.(2015)]{2015RAA....15..860Y} Yi, Z.-P., Luo, A.-L., Zhao, J.-K., et al.\ 2015, Research in Astronomy and Astrophysics, 15, 860 
\bibitem[Yuan et al.(2013)]{2013MNRAS.430.2188Y} Yuan, H.~B., Liu, X.~W., \& Xiang, M.~S.\ 2013, \mnras, 430, 2188 
\bibitem[York et al.(2000)]{2000AJ....120.1579Y} York, D.~G., Adelman, J., Anderson, J.~E., Jr., et al.\ 2000, \aj, 120, 1579
\bibitem[Zhang et al.(2016)]{2016NewA...44...66Z} Zhang, L., Pi, Q., Han, X.~L., et al.\ 2016, \na, 44, 66 
\bibitem[Zhang et al.(2017)]{2017MNRAS.464.3040Z} Zhang, Z.~H., Pinfield, D.~J., G{\'a}lvez-Ortiz, M.~C., et al.\ 2017, \mnras, 464, 3040 
\end{thebibliography}
\end{document}